%% file: BERT4Rec.tex
\documentclass[sigconf]{acmart}

\usepackage{balance}

\def\BibTeX{{\rm B\kern-.05em{\sc i\kern-.025em b}\kern-.08emT\kern-.1667em\lower.7ex\hbox{E}\kern-.125emX}}
    
\usepackage{booktabs} %
\usepackage{multirow}

\usepackage[english]{babel}
\usepackage[inline]{enumitem}

\usepackage{amsmath}
\usepackage{amsfonts}
\usepackage{bbm}
\usepackage{bm}
\usepackage{scalerel} %

\usepackage{subcaption}
\usepackage{adjustbox}

\usepackage{tikz}
\usetikzlibrary{arrows.meta}
\usetikzlibrary{backgrounds}
\usetikzlibrary{positioning, fit, mindmap, trees, calc,tikzmark,shapes}
\usetikzlibrary{shapes.arrows, fadings, automata,tikzmark,decorations.pathreplacing,patterns}
\usepackage{pgfplots}
\usepgfplotslibrary{groupplots}
\pgfplotsset{compat=1.14}
\usepackage{tikzpeople}

\usepackage{graphicx}

\usepackage{sistyle}
\SIthousandsep{,}

\input{color}

\renewcommand{\arraystretch}{0.98}

\makeatletter
\g@addto@macro\normalsize{%
  \abovedisplayskip 4pt plus 2pt minus 3pt%
  \belowdisplayskip \abovedisplayskip
  \abovedisplayshortskip 4pt plus2pt  minus3pt%
  \belowdisplayshortskip 4pt plus2pt minus3pt%
}

\makeatother

\setcopyright{acmlicensed}

\begin{document}

\copyrightyear{2019} 
\acmYear{2019} 
\acmConference[CIKM '19]{The 28th ACM International Conference on Information and Knowledge Management}{November 3--7, 2019}{Beijing, China}
\acmBooktitle{The 28th ACM International Conference on Information and Knowledge Management (CIKM '19), November 3--7, 2019, Beijing, China}
\acmPrice{15.00}
\acmDOI{10.1145/3357384.3357895}
\acmISBN{978-1-4503-6976-3/19/11}

\fancyhead{}

\title[BERT4Rec]{BERT4Rec: Sequential Recommendation with Bidirectional Encoder Representations from Transformer}

\author{Fei Sun, Jun Liu, Jian Wu, Changhua Pei, Xiao Lin, Wenwu Ou, and Peng Jiang}
\affiliation{%
  \institution{Alibaba Group, Beijing, China}
  \city{}
  \country{}
}
\email{{ofey.sf, yanhan.lj, joshuawu.wujian, changhua.pch, hc.lx, santong.oww, jiangpeng.jp}@alibaba-inc.com}

\begin{abstract}

Modeling users' dynamic preferences from their historical behaviors is challenging and crucial for recommendation systems. 
Previous methods employ sequential neural networks to encode users' historical interactions from left to right into hidden representations for making recommendations.
Despite their effectiveness, we argue that such left-to-right unidirectional models are sub-optimal due to the limitations including:
\begin {enumerate*}[label=\bfseries\itshape\alph*\upshape)]
\item unidirectional architectures restrict the power of hidden representation in users' behavior sequences;
\item they often assume a rigidly ordered sequence which is not always practical.
\end {enumerate*}
To address these limitations, we proposed a sequential recommendation model called \textbf{BERT4Rec}, which employs the deep bidirectional self-attention to model user behavior sequences.
To avoid the information leakage and efficiently train the bidirectional model, we adopt the Cloze objective to sequential recommendation, predicting the random masked items in the sequence by jointly conditioning on their left and right context.
In this way, we learn a bidirectional representation model to make recommendations by allowing each item in user historical behaviors to fuse information from both left and right sides.
Extensive experiments on four benchmark datasets show that our model outperforms various state-of-the-art sequential models consistently.

\end{abstract}



\maketitle

\input{intro}

\input{related}

\input{model}

\input{experiment}

\section{Conclusion and Future Work}

Deep bidirectional self-attention architecture has achieved tremendous success in language understanding.
In this paper, we introduce a deep bidirectional sequential model called BERT4Rec for sequential recommendation.
For model training, we introduce the Cloze task which predicts the masked items using both left and right context.
Extensive experimental results on four real-world datasets show that our model outperforms state-of-the-art baselines.

Several directions remain to be explored.
A valuable direction is to incorporate rich item features (\textit{e.g.}, category and price for products, cast for movies) into BERT4Rec instead of just modeling item ids. %
Another interesting direction for the future work would be introducing user component into the model for explicit user modeling when the users have multiple sessions.

\bibliographystyle{ACM-Reference-Format}
\bibliography{BERT4Rec}

\end{document}

%% file: color.tex
\usepackage{xcolor}

\definecolor{riptide}{RGB}{141,211,199}
\definecolor{pale_prim}{RGB}{255,255,179}
\definecolor{lavender_gray}{RGB}{190,186,218}
\definecolor{salmon}{RGB}{242,131,107}
\definecolor{seagull}{RGB}{128,177,211}
\definecolor{rajah}{RGB}{253,180,98}
\definecolor{yellow_green}{RGB}{198,222,119}
\definecolor{classic_rose}{RGB}{252,205,229}
\definecolor{feijoa}{RGB}{178,223,138}

\definecolor{cruise}{RGB}{179,226,205}
\definecolor{apricot}{RGB}{253,205,172}
\definecolor{periwinkle}{RGB}{203,213,232}
\definecolor{snow_flurry}{RGB}{230,245,201}
\definecolor{buttermilk}{RGB}{255,242,174}

\definecolor{sundown}{RGB}{249, 180, 181}
\definecolor{spindle}{RGB}{179,205,227}
\definecolor{tea_green}{RGB}{204,235,197}
\definecolor{languid_lavender}{RGB}{222,203,228}
\definecolor{champagne}{RGB}{254,217,166}
\definecolor{cream}{RGB}{255,255,204}

\definecolor{monte_carlo}{RGB}{135,204,194}
\definecolor{melon}{RGB}{254,191,181}
\definecolor{granny_smith_apple}{RGB}{150,214,150}
\definecolor{watusi}{RGB}{254,221,207}
\definecolor{see_green}{RGB}{161,228,195}

\definecolor{moss_green}{RGB}{170,216,176}
\definecolor{opal}{RGB}{164,207,190}

\definecolor{pale_turquoise}{RGB}{172,240,242}
\definecolor{Madang}{RGB}{190,235,159}
\definecolor{pixie_green}{RGB}{183,214,170}
\definecolor{coral_andy}{RGB}{243,204,205}
\definecolor{manhattan}{RGB}{226,180,125}
\definecolor{quartz}{RGB}{219,223,238}
\definecolor{spring_sun}{RGB}{242,243,195}
\definecolor{dairy_cream}{RGB}{254,226,189}
\definecolor{surf_crest}{RGB}{205,230,208}
\definecolor{french_pass}{RGB}{195,232,246}
\definecolor{cosmos}{RGB}{248,209,210}
\definecolor{portafino}{RGB}{245,237,160}
\definecolor{sail}{RGB}{163,205,235}
\definecolor{hint_green}{RGB}{226,246,209}

\definecolor{jet_stream}{RGB}{188, 214, 210}

\definecolor{azalea}{RGB}{251, 196, 196}
\definecolor{wewak}{RGB}{244, 143, 150}
\definecolor{bittersweet}{RGB}{255,111,105}
\definecolor{sunset_orange}{RGB}{242,89,75}
\definecolor{light_coral}{RGB}{244, 127, 123}
\definecolor{carnation}{RGB}{245, 80, 86}
\definecolor{flamingo}{RGB}{237, 88, 85}
\definecolor{carmine_pink}{RGB}{231, 76, 60}
\definecolor{deep_carmine_pink}{RGB}{236, 50, 67}
\definecolor{fire_engine_red}{RGB}{210,44,41}
\definecolor{amaranth}{RGB}{234,46,73}
\definecolor{ku_crimson}{RGB}{243, 0, 25}
\definecolor{fire_engine_red}{RGB}{206, 37, 51}
\definecolor{copper_rust}{RGB}{155, 64, 74}

\definecolor{chilean_fire}{RGB}{215, 87, 44}

\definecolor{japanese_laurel}{RGB}{53, 116, 40}

\definecolor{turmeric}{RGB}{211, 178, 76}
\definecolor{saffron}{RGB}{249,193,62}
\definecolor{my_sin}{RGB}{255, 176, 59}
\definecolor{tree_poppy}{RGB}{246, 154, 27}
\definecolor{jaffa}{RGB}{240, 131, 58}
\definecolor{crusta}{RGB}{254, 127, 44}
\definecolor{tahiti_gold}{RGB}{223, 102, 36}
\definecolor{outrageous_orange}{RGB}{255, 100, 45}
\definecolor{safety_orange}{RGB}{254, 106, 0}

\definecolor{turquoise}{RGB}{41,217,194}
\definecolor{puerto_rico}{RGB}{94, 194, 166}
\definecolor{mountain_meadow}{RGB}{0, 163, 136}
\definecolor{free_speech_aquamarine}{RGB}{0, 156, 114}
\definecolor{java}{RGB}{2,190,196}

\definecolor{matisse}{RGB}{25, 104, 167}
\definecolor{shakespeare}{RGB}{85, 154, 193}
\definecolor{mona_lisa}{RGB}{246,152,134}

\definecolor{bgc}{RGB}{245,245,245}
\definecolor{tuatara}{RGB}{67, 67, 67}
\definecolor{aluminum}{RGB}{153,153,153}
\definecolor{silver}{RGB}{191,191,191}
\definecolor{platinum}{RGB}{228,228,228}
\definecolor{mercury}{RGB}{230,230,230}
\definecolor{gallery}{RGB}{240,240,240}
\definecolor{athens_gray}{RGB}{236, 240, 241}
\definecolor{ship_gray}{RGB}{77,77,77}

\definecolor{early_dawn}{RGB}{252,243,218}
\definecolor{egg_shell}{RGB}{238, 234, 215}
\definecolor{midnight}{RGB}{0, 29, 50}
\definecolor{sundown}{RGB}{249, 180, 181}
\definecolor{sun_shade}{RGB}{255, 144, 68}
\definecolor{sushi}{RGB}{117, 168, 47}
\definecolor{tomato}{RGB}{255, 97, 56}
\definecolor{ice_cold}{RGB}{169,232,220}

\definecolor{jelly_bean}{RGB}{45, 126, 150}
\definecolor{celestial_blue}{RGB}{52, 152, 219}
\definecolor{curious_blue}{RGB}{41, 128, 185}
\definecolor{french_blue}{RGB}{0, 112, 182}
\definecolor{matisse}{RGB}{25, 104, 167}

\definecolor{biscay}{RGB}{44, 62, 80}

\definecolor{cosmic_latte}{RGB}{222, 247, 229}
\definecolor{chinook}{RGB}{163, 232, 178}
\definecolor{padua}{RGB}{121, 189, 143}
\definecolor{ocean_green}{RGB}{79, 176, 112}
\definecolor{pastel_green}{RGB}{107, 227, 135}
\definecolor{chateau_green}{RGB}{69, 191, 85}
\definecolor{RoyalBlue}{RGB}{69, 191, 85}
\definecolor{pigment_green}{RGB}{0, 175, 79}
\definecolor{fern}{RGB}{101,197,117}
\definecolor{killarney}{RGB}{56, 113, 66}
\definecolor{viridian}{RGB}{70, 137, 102}

%% file: intro.tex
\section{Introduction}

Accurately characterizing users' interests lives at the heart of an effective recommendation system.
In many real-world applications, users' current interests are intrinsically dynamic and evolving, influenced by their historical behaviors.
For example, one may purchase accessories (\textit{e.g.}, Joy-Con controllers) soon after buying a Nintendo Switch, though she/he will not buy console accessories under normal circumstances.

To model such sequential dynamics in user behaviors, various methods have been proposed to make \textit{sequential recommendations} based on users' historical interactions~\cite{Rendle:WWW2010:FPM,Hidasi:ICLR2016:gru4rec,Kang:ICDM2018:SAS}.
They aim to predict the successive item(s) that a user is likely to interact with given her/his past interactions.
Recently, a surge of works employ sequential neural networks, \textit{e.g.}, Recurrent Neural Network (RNN), for sequential recommendation and obtain promising results~\cite{Hidasi:ICLR2016:gru4rec,Yu:sigir2016:DRM,Wu:wsdm2017:RRN,Donkers:recsys2017:SUR,Hidasi:CIKM2018:RNN}.
The basic paradigm of previous work is to encode a user's historical interactions into a vector (\textit{i.e.}, representation of user's preference) using a left-to-right sequential model and make recommendations based on this hidden representation.

Despite their prevalence and effectiveness, we argue that such left-to-right unidirectional models are not sufficient to learn optimal representations for user behavior sequences. %
The major limitation, as illustrated in Figure~\ref{fig:model}c and~\ref{fig:model}d, is that such unidirectional models restrict the power of hidden representation for items in the historical sequences, where each item can only encode the information from previous items.
Another limitation is that previous unidirectional models are originally introduced for sequential data with natural order, \textit{e.g.}, text and time series data.
They often assume a rigidly ordered sequence over data which is not always true for user behaviors in real-world applications.
In fact, the choices of items in a user's historical interactions may not follow a rigid order assumption~\cite{Hu:ijcai2017:Diversifying,Wang:aaai2018:atem} due to various unobservable external factors~\cite{Covington:recsys2016:DNN}. %
In such a situation, it is crucial to incorporate context from both directions in user behavior sequence modeling.

To address the limitations mentioned above, we seek to use a bidirectional model to learn the representations for users' historical behavior sequences.
Specifically, inspired by the success of BERT~\cite{Devlin:2018:BERT} in text understanding, we propose to apply the deep bidirectional self-attention model to sequential recommendation, as illustrated in Figure~\ref{fig:model}b. 
For representation power, the superior results for deep bidirectional models on text sequence modeling tasks show that it is beneficial to incorporate context from both sides for sequence representations learning~\cite{Devlin:2018:BERT}.
For rigid order assumption, our model is more suitable than unidirectional models in modeling user behavior sequences since all items in the bidirectional model can leverage the contexts from both left and right side.

However, it is not straightforward and intuitive to train the bidirectional model for sequential recommendation. %
Conventional sequential recommendation models are usually trained left-to-right by predicting the next item for each position in the input sequence.
As shown in Figure~\ref{fig:model}, jointly conditioning on both left and right context in a deep bidirectional model would cause information leakage, \textit{i.e.}, allowing each item to indirectly ``see the target item''.
This could make predicting the future become trivial and the network would not learn anything useful.

To tackle this problem, we introduce the Cloze task~\cite{Devlin:2018:BERT,Wilson:Cloze} to take the place of the objective in unidirectional models (\textit{i.e.}, sequentially predicting the next item).
Specifically, we randomly mask some items (\textit{i.e.}, replace them with a special token \texttt{[mask]}) in the input sequences, and then predict the ids of those masked items based on their surrounding context. %
In this way, we avoid the information leakage and learn a bidirectional representation model by allowing the representation of each item in the input sequence to fuse both the left and right context.
In addition to training a bidirectional model, another advantage of the Cloze objective is that it can produce more samples to train a more powerful model in multiple epochs. %
However, a downside of the Cloze task is that it is not consistent with the final task (\textit{i.e.}, sequential recommendation).
To fix this, during the test, we append the special token ``\texttt{[mask]}'' at the end of the input sequence to indicate the item that we need to predict, and then make recommendations base on its final hidden vector.
Extensive experiments on four datasets show that our model outperforms various state-of-the-art baselines consistently.

The contributions of our paper are as follows:
\begin{itemize}%
    \item We propose to model user behavior sequences with a bidirectional self-attention network through Cloze task. To the best of our knowledge, this is the first study to introduce deep bidirectional sequential model and Cloze objective into the field of recommendation systems.
    \item We compare our model with state-of-the-art methods and demonstrate the effectiveness of both bidirectional architecture and the Cloze objective through quantitative analysis on four benchmark datasets.
    \item We conduct a comprehensive ablation study to analyze the contributions of key components in the proposed model. %
\end{itemize}

%% file: related.tex
\section{Related Work}

In this section, we will briefly review several lines of works closely related to ours, including general recommendation, sequential recommendation, and attention mechanism.

\subsection{General Recommendation}

Early works on recommendation systems typically use Collaborative  Filtering (CF) to model users' preferences based on their  interaction histories~\cite{Sarwar:www2001:ICF,Koren:ACF:2011}.
Among various CF methods, Matrix Factorization (MF) is the most popular one, which projects users and items into a shared vector space and estimate a user's preference on an item by the inner product between their vectors~\cite{Salakhutdinov:nips2007:PMF,Koren:2009:MFT,Koren:ACF:2011}.
Another line of work is item-based neighborhood methods \cite{Sarwar:www2001:ICF,Linden:ieee2003:ARI,Koren:kdd2008:FMN,Kabbur:kdd2013:FFI}.
They estimate a user's preference on an item via measuring its similarities with the items in her/his interaction history using a precomputed item-to-item similarity matrix.

Recently, deep learning has been revolutionizing the recommendation systems dramatically.
The early pioneer work is a two-layer Restricted Boltzmann Machines (RBM) for collaborative filtering, proposed by \citeauthor{Salakhutdinov:icml2007:RBM}~\cite{Salakhutdinov:icml2007:RBM} in Netflix Prize\footnote{\url{https://www.netflixprize.com}}. %

 One line of deep learning based methods seeks to improve the recommendation performance by integrating the distributed item representations learned from auxiliary information, \textit{e.g.}, text~\cite{Wang:kdd2015:CDL,Kim:recsys2016:CMF}, images~\cite{Wang:www2017:YIR,Kang:icdm2017:Visually}, and acoustic features~\cite{Oord:nips2013:Deep} into CF models. 
Another line of work seeks to take the place of conventional matrix factorization. 
For example, Neural Collaborative Filtering (NCF)~\cite{He:www2017:NCF} estimates user preferences via Multi-Layer Perceptions (MLP) instead of inner product, while AutoRec~\cite{Sedhain:www2015:AAM} and CDAE~\cite{Wu:wsdm2016:CDA} predict users' ratings using Auto-encoder framework.

\subsection{Sequential Recommendation}

Unfortunately, none of the above methods is for sequential recommendation since they all ignore the order in users' behaviors.

Early works on sequential recommendation usually capture sequential patterns from user historical interactions using Markov chains (MCs).
For example, \citeauthor{Shani:jmlr2005:MRS}~\cite{Shani:jmlr2005:MRS} formalized recommendation generation as a sequential optimization problem and employ Markov Decision Processes (MDPs) to address it. 
Later, \citeauthor{Rendle:WWW2010:FPM}~\cite{Rendle:WWW2010:FPM} combine the power of MCs and MF to model both sequential behaviors and general interests by Factorizing Personalized Markov Chains (FPMC).
Besides the first-order MCs, high-order MCs are also adopted to  consider more previous items~\cite{He:icdm16:Fusing,He:recsys2017:TR}. 
 
Recently, RNN and its variants, Gated Recurrent Unit (GRU)~\cite{Cho:emnlp2014:gru} and Long Short-Term Memory (LSTM)~\cite{Hochreiter:LSTM:NC1997}, are becoming more and more popular for modeling user behavior sequences~\cite{Hidasi:ICLR2016:gru4rec,Yu:sigir2016:DRM,Wu:wsdm2017:RRN,Donkers:recsys2017:SUR,Quadrana:recsys2017:PSR,Li:cikm2017:NAS,Hidasi:CIKM2018:RNN}.
The basic idea of these methods is to encode user's previous records into a vector (\textit{i.e.}, representation of user's preference which is used to make predictions) with various recurrent architectures and loss functions, including session-based GRU with ranking loss (GRU4Rec)~\cite{Hidasi:ICLR2016:gru4rec}, Dynamic REcurrent bAsket Model (DREAM)~\cite{Yu:sigir2016:DRM}, user-based GRU~\cite{Donkers:recsys2017:SUR}, attention-based GRU (NARM)~\cite{Li:cikm2017:NAS}, and improved GRU4Rec with new loss function (\textit{i.e.}, \texttt{BPR-max} and \texttt{TOP1-max}) and an improved sampling strategy~\cite{Hidasi:CIKM2018:RNN}.

Other than recurrent neural networks, various deep learning models are also introduced for sequential recommendation~\cite{Chen:wsdm2018:SRU,Tang:WSDM2018:caser,Liu:kdd2018:STAMP,Kang:ICDM2018:SAS}. %
For example, \citeauthor{Tang:WSDM2018:caser}~\cite{Tang:WSDM2018:caser} propose a Convolutional Sequence  Model (Caser) to learn sequential patterns using both horizontal and vertical convolutional filters.
\citeauthor{Chen:wsdm2018:SRU}~\cite{Chen:wsdm2018:SRU} and \citeauthor{Huang:sigir2018:ISR}~\cite{Huang:sigir2018:ISR} employ Memory Network to improve sequential recommendation.
STAMP captures both users' general interests and current interests using an MLP network with attention~\cite{Liu:kdd2018:STAMP}.

\subsection{Attention Mechanism}

Attention mechanism has shown promising potential in modeling sequential data, \textit{e.g.}, machine translation~\cite{Bahdanau:iclr2015,Vaswani:nips2017:Attention} and text classification~\cite{Yang:naacl2016:HAN}. 
Recently, some works try to employ the attention mechanism to improve recommendation performances and interpretability~\cite{Li:cikm2017:NAS,Liu:kdd2018:STAMP}.
For example, \citeauthor{Li:cikm2017:NAS}~\cite{Li:cikm2017:NAS} incorporate an attention mechanism into GRU to capture both the user's sequential behavior and main purpose in session-based recommendation.

\input{model_fig}

The works mentioned above basically treat attention mechanism as an additional component to the original models. %
In contrast, Transformer~\cite{Vaswani:nips2017:Attention} and BERT~\cite{Devlin:2018:BERT} are built solely on multi-head self-attention and achieve state-of-the-art results on text sequence modeling.
Recently, there is a rising enthusiasm for applying purely attention-based neural networks to model sequential data for their effectiveness and efficiency~\cite{Lin:iclr2017,Shaw:naacl2018:self,Liu:iclr2018:Generating,Radford:2018:GPT,Radford:2019:Language}.
For sequential recommendation, \citeauthor{Kang:ICDM2018:SAS}~\cite{Kang:ICDM2018:SAS} introduce a two-layer \textit{Transformer decoder} (\textit{i.e.}, Transformer language model) called SASRec to capture user's sequential behaviors and achieve state-of-the-art results on several public datasets.
SASRec is closely related to our work.
However, it is still a unidirectional model using a casual attention mask.
While we use a bidirectional model to encode users' behavior sequences with the help of Cloze task.

%% file: model_fig.tex
\tikzset{
  emb/.style = {draw, rectangle, fill=cosmos, minimum width=3em, minimum height=1.3em, inner sep=0, outer sep=0},
  transformer/.style = {draw, rectangle, fill=cruise, minimum width=3em, minimum height=1.5em},
  norm/.style = {draw, rectangle, fill=spring_sun, minimum width=4.5em, minimum height=1em, inner sep=0, outer sep=0},
  dropout/.style = {draw, rectangle, fill=hint_green, minimum width=4.5em, minimum height=1em, inner sep=0, outer sep=0},
  mh/.style = {draw, rectangle, fill=dairy_cream, minimum width=4.5em, minimum height=1.5em},
  ff/.style = {draw, rectangle, fill=french_pass, minimum width=4.5em, minimum height=1.5em},
  posemb/.style = {draw, rectangle, fill=watusi, minimum width=3em, minimum height=1.3em, inner sep=0, outer sep=0},
  proj/.style = {draw, rectangle, fill=Madang, minimum width=3em, minimum height=1.5em, inner sep=0, outer sep=0},
  FARROW/.style={arrows={-{Latex[length=1mm, width=0.8mm]}}}
}

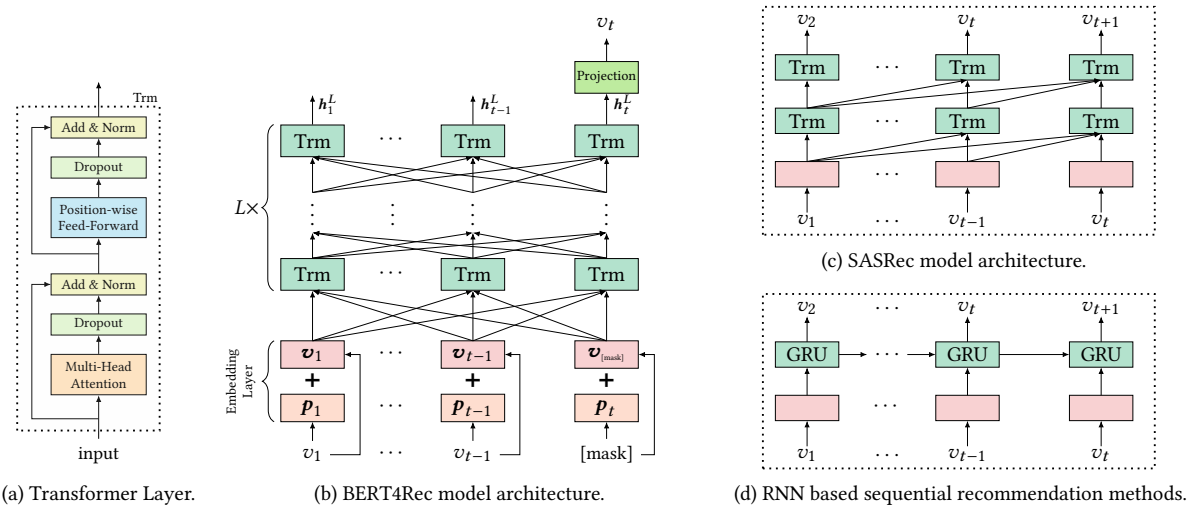
\begin{figure*}
\centering
\resizebox{0.9\textwidth}{!}{
\begin{tikzpicture}

\node[emb] (e1) at (0, 0) {$\bm{v}_1$};
\node[right of=e1, node distance=1.2cm] (e2) {$\cdots$};
\node[emb, right of=e2, node distance=1.2cm] (e3) {$\bm{v}_{t-1}$};
\node[emb, right of=e3, node distance=2cm] (e4) {$\bm{v}_{\scaleto{\text{[mask]}}{3pt}}$};

\node[below of=e1, node distance=1.5cm] (i1) {$v_1$};
\node[below of=e2, node distance=1.5cm] (i2) {$\cdots$};
\node[below of=e4, node distance=1.5cm] (i4) {\small{[mask]}};
\node[below of=e3, node distance=1.5cm] (i3) {$v_{t-1}$};

\foreach \x in {1,3,4}
{
\node[transformer, above of=e\x, node distance=1.2cm] (t\x) {Trm};
\node[above of=t\x, node distance=1cm] (d\x) {$\vdots$};
\node[transformer, above of=d\x, node distance=1cm] (tt\x) {Trm};
\node[below of=e\x, node distance=0.4cm] (add\x) {\huge\textbf{+}};
}

\draw [decorate, decoration={brace,amplitude=5pt}] ([xshift=-1mm] t1.south west) -- ([xshift=-1mm] tt1.north west) node[midway,xshift=-0.4cm, align=center] (w) {$L\times$};

\node[posemb, below of=e1, node distance=0.8cm] (pe1) {$\bm{p}_1$};
\node[below of=e2, node distance=0.8cm] (pe2) {$\cdots$};
\node[posemb, below of=e3, node distance=0.8cm] (pe3) {$\bm{p}_{t-1}$};
\node[posemb, below of=e4, node distance=0.8cm] (pe4) {$\bm{p}_{t}$};

\node[proj, above of=tt4, node distance=0.95cm] (proj1) {\scriptsize Projection};

\node[above of=tt1, node distance=0.8cm] (o1) {};
\node[above of=tt1, node distance=0.55cm, xshift=0.23cm] (oa1) {\scriptsize$\bm{h}_1^L$};
\node[above of=tt3, node distance=0.8cm] (o3) {};
\node[above of=tt3, node distance=0.55cm, xshift=0.35cm] (oa3) {\scriptsize$\bm{h}_{t-1}^L$};
\node[above of=proj1, node distance=0.8cm] (o4) {$v_{t}$};
\node[above of=tt4, node distance=0.55cm, xshift=0.25cm] (oa4) {\scriptsize$\bm{h}_{t}^L$};

\draw[FARROW] (tt4) -> (proj1) ;
\draw[FARROW] (proj1) -> (o4) ;
\draw[FARROW] (tt1) -> (o1) ;
\draw[FARROW] (tt3) -> (o3) ;

\draw[FARROW] (i1.east) -- ++(0.45, 0) |-  (e1.east) ;
\draw[FARROW] (i3.east) -- ++(0.3, 0) |-  (e3.east) ;
\draw[FARROW] (i4.east) -- ++(0.2, 0) |-  (e4.east) ;

\foreach \x in {1,3,4}
{
\draw[FARROW] (i\x) -- (pe\x) ;
}

\draw [decorate, decoration={brace,amplitude=5pt}] ([xshift=-1mm] pe1.south west) -- ([xshift=-1mm] e1.north west) node[midway,xshift=-0.5cm, rotate=90, align=center, font=\scriptsize] (w) {Embedding\\ Layer};

\foreach \x in {1,3,4}
{
\foreach \y in {1,3,4}
{
    \draw[FARROW] (e\x.north) -> (t\y.south) ;
    \draw[FARROW] (t\x.north) -> (d\y.south) ;
    \draw[FARROW] ($(d\x.north)+(0, -0.15)$) -> (tt\y.south) ;
}
}
\node[above of=e2, node distance=1.2cm] (t2) {$\cdots$};
\node[above of=t2, node distance=1cm] (d2) {$\vdots$};
\node[above of=d2, node distance=1cm] (tt2) {$\cdots$};

\node[mh, left of=e1, node distance=3.2cm, align=center, font=\scriptsize, yshift=-0.3cm] (trm1) {Multi-Head\\ Attention};
\node[dropout, above of=trm1, node distance=0.75cm, align=center, font=\scriptsize] (do1) {Dropout};
\node[norm, above of=do1, node distance=0.6cm, align=center, font=\scriptsize] (n1) {Add \& Norm};

\draw[FARROW] (trm1) edge (do1);
\draw[FARROW] (do1) edge (n1);

\node[ff, above of=n1, node distance=1cm, align=center, font=\scriptsize] (ff1) {Position-wise\\Feed-Forward};
\node[dropout, above of=ff1, node distance=0.75cm, align=center, font=\scriptsize] (do2) {Dropout};
\node[norm, above of=do2, node distance=0.6cm, align=center, font=\scriptsize] (n2) {Add \& Norm};

\draw[FARROW] (n1) edge (ff1);
\draw[FARROW] (ff1) edge (do2);
\draw[FARROW] (do2) edge (n2);

\node[below of=trm1, node distance=1.2cm, align=center, font=\small] (input) {input};
\node[above of=n2, node distance=0.8cm, align=center, font=\scriptsize] (out) {};
\node[right of=out, node distance=0.7cm, align=center, font=\scriptsize, yshift=-0.35cm] (aa) {Trm};

\draw[FARROW] (n2) edge (out);
\draw[FARROW] (input) edge (trm1);

\draw[FARROW] ($(input.north)-(0,-0.3)$) -- ++(-1, 0) |- (n1.west) ;
\draw[FARROW] ($(n1.north)-(0,-0.3)$) -- ++(-1, 0) |- (n2.west) ;

\draw[thick,dotted, fill opacity=0.1, fill=bgc]  ($(input.south west)+(-0.8, 0.6)$) rectangle ($(out.north east)+(0.8, -0.6)$);

\node[below of=i2, node distance=0.6cm, xshift=1cm] (bert) {(b) BERT4Rec model architecture.};
\node[left of=bert, node distance=5.4cm] (trm) {(a) Transformer Layer.};

\node[emb, right of=e4, node distance=3cm, yshift=-0.8cm, minimum height=1.2em] (u1) {};

\node[right of=u1, node distance=1.2cm, minimum height=1.2em] (u2) {$\cdots$};
\node[emb, right of=u2, node distance=1.2cm, minimum height=1.2em] (u3) {};
\node[emb, right of=u3, node distance=2cm, minimum height=1.2em] (u4) {};

\node[below of=u1, node distance=0.7cm] (ui1) {$v_1$};
\node[below of=u2, node distance=0.7cm] (ui2) {$\ldots$};
\node[below of=u4, node distance=0.7cm] (ui4) {$v_{t}$};
\node[below of=u3, node distance=0.7cm] (ui3) {$v_{t-1}$};

\foreach \x in {1,3,4}
{
\node[transformer, above of=u\x, node distance=0.8cm, minimum height=1.2em, inner sep=0, outer sep=0] (ut\x) {GRU};
\draw[FARROW] (u\x) edge (ut\x) ;
}
\node[above of=u2, node distance=0.8cm] (ut2) {$\ldots$};
\foreach \x/\y in {1/2, 2/3, 3/4}
{
\draw[FARROW] (ut\x) edge (ut\y) ;
}

\node[above of=ut1, node distance=0.7cm] (utt1) {$v_2$};
\node[above of=ut2, node distance=0.7cm] (utt2) {$\ldots$};
\node[above of=ut3, node distance=0.7cm] (utt3) {$v_{t}$};
\node[above of=ut4, node distance=0.7cm] (utt4) {$v_{t+1}$};
\foreach \x/\y in {1,3,4}
{
\draw[FARROW] ($(ui\x.north)-(0,0.08)$) -> (u\x) ;
\draw[FARROW] (ut\x) -> ($(utt\x.south)-(0,-0.08)$) ;
}

\draw[thick,dotted]  ($(ui1.south west)+(-0.4, 0)$) rectangle ($(utt4.north east)+(0.4, 0)$);
\node[below of=ui2, node distance=0.6cm, xshift=1.1cm] (rnn) {(d) RNN based sequential recommendation methods.};

\foreach \x in {1,3,4}
{
\node[emb, above of=utt\x, node distance=2cm, minimum height=1.2em] (sa\x) {};
\node[transformer, above of=sa\x, node distance=0.8cm, minimum height=1.2em, inner sep=0, outer sep=0] (st\x) {Trm};
\node[transformer, above of=st\x, node distance=0.8cm, minimum height=1.2em, inner sep=0, outer sep=0] (stt\x) {Trm};
}
\node[below of=sa1, node distance=0.7cm] (si1) {$v_1$};
\node[below of=sa3, node distance=0.7cm] (si3) {$v_{t-1}$};
\node[below of=sa4, node distance=0.7cm] (si4) {$v_{t}$};
\node[right of=sa1, node distance=1.2cm, minimum height=1.2em] (sa2) {$\ldots$};

\node[below of=sa2, node distance=0.7cm, minimum height=1.2em] (si2) {$\ldots$};
\node[above of=sa2, node distance=0.8cm, minimum height=1.2em] (st2) {$\ldots$};
\node[above of=st2, node distance=0.8cm, minimum height=1.2em] (stt2) {$\ldots$};
\node[above of=stt1, node distance=0.7cm] (so1) {$v_2$};
\node[above of=stt3, node distance=0.7cm] (so3) {$v_{t}$};
\node[above of=stt4, node distance=0.7cm] (so4) {$v_{t+1}$};

\foreach \x in {1,3,4}
{
\draw[FARROW] ($(si\x.north)-(0,0.08)$) -> (sa\x) ;
\draw[FARROW] (sa\x.north) -> (st\x.south) ;
\draw[FARROW] (st\x.north) -> (stt\x.south) ;
\draw[FARROW] (stt\x) -> ($(so\x.south)-(0,-0.08)$) ;
}

\foreach \x/\y in {1/3, 1/4, 3/4}
{
\draw[FARROW] (sa\x.north) -> (st\y.south) ;
\draw[FARROW] (st\x.north) -> (stt\y.south) ;
}

\draw[thick,dotted]  ($(si1.south west)+(-0.4, 0)$) rectangle ($(so4.north east)+(0.4, 0)$);
\node[below of=si2, node distance=0.6cm, xshift=1cm] (sas) {(c) SASRec model architecture.};

\end{tikzpicture}}
    \caption{Differences in sequential recommendation model architectures. BERT4Rec learns a bidirectional model via Cloze task, while SASRec and RNN based methods are all left-to-right unidirectional model which predict next item sequentially.}
    \label{fig:model}
\end{figure*}

%% file: model.tex
\section{BERT4Rec}
Before going into the details, we first introduce the research problem, the basic concepts, and the notations in this paper.

\subsection{Problem Statement}
In sequential recommendation, let $\mathcal{U}{=}\{u_1,u_2,\dots,u_{\vert\mathcal{U}\vert}\}$ denote a set of users, $\mathcal{V}{=}\{v_1,v_2,\dots,v_{|\mathcal{V}|}\}$ be a set of items, and list $\mathcal{S}_u{=}[v_1^{(u)},\dots,v_t^{(u)},\dots,v_{n_u}^{(u)}]$ denote the interaction sequence in chronological order for user $u \in \mathcal{U}$, where $v_t^{(u)} \in \mathcal{V}$ is the item that $u$ has interacted with at time step\footnote{Here, following~\cite{Rendle:WWW2010:FPM,Kang:ICDM2018:SAS}, we use the relative time index instead of absolute time index for numbering interaction records.} $t$ and $n_u$ is the the length of interaction sequence for user $u$.
Given the interaction history $\mathcal{S}_u$, sequential recommendation aims to predict the item that user $u$ will interact with at time step $n_{u} + 1$.
It can be formalized as modeling the probability over all possible items for user $u$ at time step $n_{u} {+} 1$:
\begin{equation*}
    p\bigl(v_{n_u+1}^{(u)}=v|\ \mathcal{S}_u\bigr)
\end{equation*}

\subsection{Model Architecture}

Here, we introduce a new sequential recommendation model called \textbf{BERT4Rec}, which adopts \textbf{B}idirectional \textbf{E}ncoder \textbf{R}epresentations from \textbf{T}ransformers to a new task, sequential \textbf{R}ecommendation.
It is built upon the popular self-attention layer, ``Transformer layer''. %

As illustrated in Figure~\ref{fig:model}b, BERT4Rec is stacked by $L$ bidirectional Transformer layers.
At each layer, it iteratively revises the representation of every position by exchanging information across all positions at the previous layer in parallel with the Transformer layer.
Instead of learning to pass relevant information forward step by step as RNN based methods did in Figure~\ref{fig:model}d, self-attention mechanism endows BERT4Rec with the capability to directly capture the dependencies in any distances.
This mechanism results in a global receptive field, while CNN based methods like Caser usually have a limited receptive field.
In addition, in contrast to RNN based methods, self-attention is straightforward to parallelize.

Comparing Figure~\ref{fig:model}b,~\ref{fig:model}c, and~\ref{fig:model}d, the most noticeable difference is that SASRec and RNN based methods are all left-to-right unidirectional architecture, while our BERT4Rec uses bidirectional self-attention to model users' behavior sequences.
In this way, our proposed model can obtain more powerful representations of users' behavior sequences to improve  recommendation performances.

\subsection{Transformer Layer}

As illustrated in Figure~\ref{fig:model}b, given an input sequence of length $t$, %
we iteratively compute hidden representations $\bm{h}_i^{l}$ at each layer $l$ for each position $i$ simultaneously by applying the Transformer layer from~\cite{Vaswani:nips2017:Attention}.
Here, we stack $\bm{h}_i^{l} \in \mathbb{R}^d$ together into matrix $\bm{H}^{l} {\in} \mathbb{R}^{t\times d}$ since we compute attention function on all positions simultaneously in practice. 
As shown in Figure~\ref{fig:model}a, the Transformer layer \texttt{Trm} contains two sub-layers, a \textit{Multi-Head Self-Attention} sub-layer and a \textit{Position-wise Feed-Forward Network}.

\textbf{Multi-Head Self-Attention}.
Attention mechanisms have become an integral part of sequence modeling in a variety of tasks, allowing capturing the dependencies between representation pairs without regard to their distance in the sequences.
Previous work has shown that it is beneficial to jointly attend to information from different representation subspaces at different positions~\cite{Vaswani:nips2017:Attention,Li:emnlp2018:Multi,Devlin:2018:BERT}.
Thus, we here adopt the multi-head self-attention instead of performing a single attention function.
Specifically, multi-head attention first linearly projects $\bm{H}^l$ into $h$ subspaces, with different, learnable linear projections, and then apply $h$ attention functions in parallel to produce the output representations which are concatenated and once again projected:
\begin{equation}
    \begin{aligned}
        \mathtt{MH}(\bm{H}^{l}) &= [\text{head}_1; \text{head}_2; \dots; \text{head}_h]\bm{W}^{O} \\
        \text{head}_i &= \mathtt{Attention}\bigl(\bm{H}^l\bm{W}_i^Q, \bm{H}^l\bm{W}_i^K, \bm{H}^l\bm{W}_i^V \bigr)
    \end{aligned}
\label{eq:mh}
\end{equation} %
where the projections matrices for each head $\bm{W}_i^Q \in \mathbb{R}^{d \times d/h}$, $\bm{W}_i^K \in \mathbb{R}^{d \times d/h}$, $\bm{W}_i^V \in \mathbb{R}^{d \times d/h}$, and $\bm{W}_i^O \in \mathbb{R}^{d \times d}$ are learnable parameters. %
Here, we omit the layer subscript $l$ for the sake of simplicity.
In fact, these projection parameters are not shared across the layers.
Here, the $\mathtt{Attention}$ function is \textit{Scaled Dot-Product Attention}: %
\begin{equation}
    \mathtt{Attention}(\bm{Q}, \bm{K}, \bm{V}) = \mathrm{softmax} \biggl(\frac{\bm{Q}\bm{K}^{\top}}{\sqrt{d/h}}\biggr) \bm{V}
    \label{eq:scale_dot}
\end{equation}
where \textit{query} $\bm{Q}$, \textit{key} $\bm{K}$, and \textit{value} $\bm{V}$ are projected from the same matrix $\bm{H}^l$ with different learned projection matrices as in Equation~\ref{eq:mh}. %
The \textit{temperature} $\sqrt{d/h}$ is introduced to produce a softer attention distribution for avoiding extremely small gradients~\cite{Hinton:nips2014:Distilling,Vaswani:nips2017:Attention}.

\textbf{Position-wise Feed-Forward Network}.
As described above, the self-attention sub-layer is mainly based on linear projections.
To endow the model with nonlinearity and interactions between different dimensions, we apply a \textit{Position-wise Feed-Forward Network} to the outputs of the self-attention sub-layer, separately and identically at each position.
It consists of two affine transformations with a Gaussian Error Linear Unit (\texttt{GELU}) activation in between:
\begin{equation}
    \begin{aligned}
      \texttt{PFFN}(\bm{H}^l) & = \bigl[\mathtt{FFN}(\bm{h}_1^{l})^{\top}; \dots; \mathtt{FFN}(\bm{h}_t^{l})^{\top}\bigr]^{\top}\\
       \mathtt{FFN}(\bm{x}) &= \texttt{GELU}\bigl(\bm{x} \bm{W}^{(1)}+\bm{b}^{(1)}\bigr)\bm{W}^{(2)} + \bm{b}^{(2)}\\
       \texttt{GELU}(x) &= x \Phi(x)
    \end{aligned}
    \label{eq:ffn}
\end{equation}
where $\Phi(x)$ is the cumulative distribution function of the standard gaussian distribution, $\bm{W}^{(1)}\in \mathbb{R}^{d\times 4d}$, $\bm{W}^{(2)}\in \mathbb{R}^{4d\times d}$, $\bm{b}^{(1)}\in \mathbb{R}^{4d}$ and $\bm{b}^{(2)}\in \mathbb{R}^{d}$ are learnable parameters and shared across all positions.
We omit the layer subscript $l$ for convenience.
In fact, these parameters are different from layer to layer.
In this work, following OpenAI GPT~\cite{Radford:2018:GPT} and BERT~\cite{Devlin:2018:BERT}, we use a smoother \texttt{GELU}~\cite{Hendrycks:gleu} activation rather than the standard \texttt{ReLu} activation. %

\textbf{Stacking Transformer Layer}.
As elaborated above, we can easily capture item-item interactions across the entire user behavior sequence using self-attention mechanism.
Nevertheless, it is usually beneficial to learn more complex item transition patterns by stacking the self-attention layers.
However, the network becomes more difficult to train as it goes deeper.
Therefore, we employ a residual connection~\cite{He:cvpr2016:resnet} around each of the two sub-layers as in Figure~\ref{fig:model}a, followed by layer normalization~\cite{Lei:Layer}.
Moreover, we also apply dropout~\cite{Srivastava:jmlr15:Dropout} to the output of each sub-layer, before it is normalized. %
That is, the output of each sub-layer is $\texttt{LN}(\bm{x} + \texttt{Dropout}(\text{sublayer}(\bm{x})))$, where $\text{sublayer}(\cdot)$ is the function implemented by the sub-layer itself, $\texttt{LN}$ is the layer normalization function defined in~\cite{Lei:Layer}. %
We use $\texttt{LN}$ to normalize the inputs over all the hidden units in the same layer for stabilizing and accelerating the network training.

In summary, BERT4Rec refines the hidden representations of each layer as follows:
\begin{align}
    \bm{H}^l &= \mathtt{Trm}\bigl(\bm{H}^{l-1}\bigr), \quad \forall i \in [1,\dots, L] \\
    \mathtt{Trm}(\bm{H}^{l-1}) &= \mathtt{LN}\Bigl(\bm{A}^{l-1} +  \mathtt{Dropout}\bigl(\mathtt{PFFN}(\bm{A}^{l-1})\bigr)\Bigr) \\
    \bm{A}^{l-1} &= \mathtt{LN}\Bigr(\bm{H}^{l-1} + \mathtt{Dropout}\bigl(\mathtt{MH}(\bm{H}^{l-1})\bigr)\Bigr) 
\end{align}

\subsection{Embedding Layer}
As elaborated above, without any recurrence or convolution module, the Transformer layer $\mathtt{Trm}$ is not aware of the order of the input sequence.
In order to make use of the sequential information of the input, we inject \textit{Positional Embeddings} into the input item embeddings at the bottoms of the Transformer layer stacks.
For a given item $v_i$, its input representation $\bm{h}_i^0$ is constructed by summing the corresponding item and positional embedding:
\begin{equation*}
    \begin{aligned}
        \bm{h}_{i}^0 &= \bm{v}_i + \bm{p}_i
    \end{aligned}
\end{equation*}
where $\bm{v}_i {\in} \bm{E}$ is the $d-$dimensional embedding for item $v_i$, $\bm{p}_i {\in} \bm{P}$ is the $d-$dimensional positional embedding for position index $i$.
In this work, we use the learnable positional embeddings instead of the fixed sinusoid embeddings in~\cite{Vaswani:nips2017:Attention} for better performances.
The positional embedding matrix $\bm{P} \in \mathbb{R}^{N\times d}$ allows our model to identify which portion of the input it is dealing with.
However, it also imposes a restriction on the maximum sentence length $N$ that our model can handle.
Thus, we need to truncate the the input sequence $[v_1,\dots,v_t]$ to the last $N$ items $[v_{t-N+1}^{u},\dots,v_t]$ if $t>N$.

\subsection{Output Layer}

After $L$ layers that hierarchically exchange information across all positions in the previous layer, we get the final output $\bm{H}^L$ for all items of the input sequence.
Assuming that we mask the item $v_t$ at time step $t$, we then predict the masked items $v_t$ base on $\bm{h}_t^L$ as shown in Figure~\ref{fig:model}b.
Specifically, we apply a two-layer feed-forward network with \texttt{GELU} activation in between to produce an output distribution over target items:
\begin{equation}
    P(v) = \mathrm{softmax}\bigl(\mathtt{GELU}(\bm{h}_t^L \bm{W}^P +\bm{b}^P) \bm{E}^{\top} + \bm{b}^O \bigr)
    \label{eq:prob}
\end{equation}
where $\bm{W}^P$ is the learnable projection matrix, $\bm{b}^P$, and $\bm{b}^O$ are bias terms, $\bm{E}\in \mathbb{R}^{|\mathcal{V}|\times d}$ is the embedding matrix for the item set $\mathcal{V}$.
We use the shared item embedding matrix in the input and output layer for alleviating overfitting and reducing model size.

\subsection{Model Learning}
\label{sec:ml}

\textbf{Training}. 
Conventional unidirectional sequential recommendation models usually train the model by predicting the next item for each position in the input sequence as illustrated in Figure~\ref{fig:model}c and~\ref{fig:model}d.
Specifically, the target of the input sequence $[v_1,\dots,v_{t}]$ is a shifted version $[v_2,\dots,v_{t+1}]$.
However, as shown in Figure~\ref{fig:model}b, jointly conditioning on both left and right context in a bidirectional model would cause the final output representation of each item to contain the information of the target item.
This makes predicting the future become trivial and the network would not learn anything useful.
A simple solution for this issue is to create $t-1$ samples (subsequences with next items like ([$v_1$], $v_2$) and ([$v_1,v_2$], $v_3$)) from the original length $t$ behavior sequence and then encode each historical subsequence with the bidirectional model to predict the target item. 
However, this approach is very time and resources consuming since we need to create a new sample for each position in the sequence and predict them separately.

In order to efficiently train our proposed model, we apply a new objective: \textit{Cloze} task~\cite{Wilson:Cloze} (also known as ``Masked Language Model'' in~\cite{Devlin:2018:BERT}) to sequential recommendation.
It is a test consisting of a portion of language with some words removed, where the participant is asked to fill the missing words.
In our case, for each training step, we randomly mask $\rho$ proportion of all items in the input sequence (\textit{i.e.}, replace with special token ``\texttt{[mask]}''), and then predict the original ids of the masked items based solely on its left and right context.
For example: 
\begin{figure}[h]
\centering
\resizebox{0.97\linewidth}{!}{
\begin{tikzpicture}[decoration={random steps,segment length=3mm, amplitude=0.5pt}]
\node[] (e1) at (0, 0) {\textbf{Input}: $[v_1,v_2,v_3,v_4,v_5]$};
\node[right of=e1, node distance=5.1cm] (e2)  {$[v_1,$ {\small$\mathtt{[mask]}_1$}$, v_3,$ {\small$\mathtt{[mask]}_2$}$, v_5]$};
\node[below of=e1, node distance=0.5cm, xshift=0.68cm] (e3)  {\textbf{Labels}: {\small$\mathtt{[mask]}_1$} $= v_2$, ~~ {\small$\mathtt{[mask]}_2$} $= v_4$};
\draw[FARROW] (e1) -> (e2) ;
\node[right of=e1, node distance=2.45cm, yshift=0.1cm] (e4)  {\scriptsize randomly mask};

\draw[decorate]  ($(e3.south west)+(-0.1, -0.1)$) rectangle ($(e2.north east)+(0.1, 0.1)$);

\end{tikzpicture}}
\end{figure}

\noindent The final hidden vectors corresponding to ``\texttt{[mask]}'' are fed into an output softmax over the item set, as in conventional sequential recommendation.
Eventually, we define the loss for each masked input $\mathcal{S}_u^{\prime}$ as the negative log-likelihood of the masked targets: %
\begin{equation}
    \mathcal{L} = \frac{1}{|\mathcal{S}_u^m|} \sum_{v_m\in \mathcal{S}_u^m} -\log P(v_m=v_m^{*}|\mathcal{S}_u^{\prime} )
    \label{eq:loss}
\end{equation}
where $\mathcal{S}_u^{\prime}$ is the masked version for user behavior history $\mathcal{S}_u$, $\mathcal{S}_u^m$ is the random masked items in it, $v_m^*$ is the true item for the masked item $v_m$, and the probability $P(\cdot)$ is defined in Equation~\ref{eq:prob}.

An additional advantage for Cloze task is that it can generate more samples to train the model.
Assuming a sequence of length $n$, conventional sequential predictions in Figure~\ref{fig:model}c and~\ref{fig:model}d  produce $n$ unique samples for training, while BERT4Rec can obtain $\binom{n}{k}$ samples (if we randomly mask $k$ items) in multiple epochs.
It allows us to train a more powerful bidirectional representation model.

\textbf{Test}.
As described above, we create a mismatch between the training and the final sequential recommendation task since the Cloze objective is to predict the current masked items while sequential recommendation aims to predict the future.
To address this, we append the special token ``\texttt{[mask]}'' to the end of user's behavior sequence, and then predict the next item based on the final hidden representation of this token.
To better match the sequential recommendation task (\textit{i.e.}, predict the last item), we also produce samples that only mask the last item in the input sequences during training.
It works like fine-tuning for sequential recommendation and can further improve the recommendation performances.

\subsection{Discussion}

Here, we discuss the relation of our model with previous related work.

\textbf{SASRec}. Obviously, SASRec is a left-to-right unidirectional version of our BERT4Rec with single head attention and causal attention mask.
Different architectures lead to different training methods.
SASRec predicts the next item for each position in a sequence, while BERT4Rec predicts the masked items in the sequence using Cloze objective.

\textbf{CBOW \& SG}. Another very similar work is Continuous Bag-of-Words (CBOW) and Skip-Gram (SG)~\cite{Mikolov:cbow}.
CBOW predicts a target word using the average of all the word vectors in its context (both left and right).
It can be seen as a simplified case of BERT4Rec, if we use one self-attention layer in BERT4Rec with uniform attention weights on items, unshare item embeddings, remove the positional embedding, and only mask the central item.
Similar to CBOW, SG can also be seen as a simplified case of BERT4Rec following similar reduction operations (mask all items except only one).
From this point of view, Cloze can be seen as a general form for the objective of CBOW and SG.
Besides, CBOW uses a simple aggregator to model word sequences since its goal is to learn good word representations, not sentence representations.
On the contrary, we seek to learn a powerful behavior sequence representation model (deep self-attention network in this work) for making recommendations.

\textbf{BERT}. Although our BERT4Rec is inspired by the BERT in NLP, it still has several differences from BERT:
\begin {enumerate*}[label=\bfseries\itshape\alph*\upshape)]
\item The most critical difference is that BERT4Rec is an end-to-end model for sequential recommendation, while BERT is a pre-training model for sentence representation.
BERT leverages large-scale task-independent corpora to pre-train the sentence representation model for various text sequence tasks since these tasks share the same background knowledge about the language.
However, this assumption does not hold in the recommendation tasks.
Thus we train BERT4Rec end-to-end for different sequential recommendation datasets.
\item Different from BERT, we remove the next sentence loss and segment embeddings since BERT4Rec models a user's historical behaviors as only one sequence in sequential recommendation task.
\end {enumerate*}

%% file: experiment.tex
\section{Experiments}

\subsection{Datasets}

\begin{table}
    \centering
    \caption{Statistics of datasets.}
    \label{tab:dataset}
    \begin{adjustbox}{max width=\linewidth}
    \begin{tabular}{l r r r r r}
    \toprule
    Datasets & \#users & \#items & \#actions & Avg. length &  Density\\
    \midrule
    Beauty & \num{40226} & \num{54542} & 0.35m & 8.8 & 0.02\% \\
    Steam & \num{281428} & \num{13044} & 3.5m & 12.4 & 0.10\% \\
    ML-1m & \num{6040} & \num{3416} & 1.0m & 163.5 & 4.79\% \\
    ML-20m & \num{138493} & \num{26744} & 20m & 144.4 & 0.54\%\\
    \bottomrule
    \end{tabular}
    \end{adjustbox}
\end{table}

We evaluate the proposed model on four real-world representative  datasets which vary significantly in domains and sparsity.

\begin{itemize}
    \item Amazon \textbf{Beauty}\footnote{\url{http://jmcauley.ucsd.edu/data/amazon/}}: This is a series of product review datasets crawled from Amazon.com by \citeauthor{McAuley:sigir2015}~\cite{McAuley:sigir2015}. 
    They split the data into separate datasets according to the top-level product categories on Amazon.
    In this work, we adopt the ``Beauty'' category.
    \item \textbf{Steam}\footnote{\url{https://cseweb.ucsd.edu/~jmcauley/datasets.html\#steam_data}}: 
    This is a dataset collected from Steam, a large online video game distribution platform, by \citeauthor{Kang:ICDM2018:SAS}~\cite{Kang:ICDM2018:SAS}.
    \item \textbf{MovieLens}~\cite{Harper:2015:MDH}: This is a popular benchmark dataset for evaluating recommendation algorithms.
    In this work, we adopt two well-established versions, MovieLens 1m (\textbf{ML-1m})\footnote{\url{https://grouplens.org/datasets/movielens/1m/}} and MovieLens 20m (\textbf{ML-20m})\footnote{\url{https://grouplens.org/datasets/movielens/20m/}}.
\end{itemize}

For dataset preprocessing, we follow the common practice in \cite{Rendle:WWW2010:FPM,Tang:WSDM2018:caser,Kang:ICDM2018:SAS}.
For all datasets, we convert all numeric ratings or the presence of a review to implicit feedback of 1 (\textit{i.e.}, the user interacted with the item).
After that, we group the interaction records by users and build the interaction sequence for each user by sorting these interaction records according to the timestamps.
To ensure the quality of the dataset, following the common practice~\cite{Rendle:WWW2010:FPM,He:www2017:NCF,Tang:WSDM2018:caser,Kang:ICDM2018:SAS}, we keep users with at least five feedbacks.
The statistics of the processed datasets are summarized in Table~\ref{tab:dataset}.

\subsection{Task Settings \& Evaluation Metrics}

To evaluate the sequential recommendation models, we adopted the \textit{leave-one-out} evaluation (\textit{i.e.}, next item recommendation) task, which has been widely used in~\cite{He:www2017:NCF,Tang:WSDM2018:caser,Kang:ICDM2018:SAS}.
For each user, we hold out the last item of the behavior sequence as the test data, treat the item just before the last as the validation set, and utilize the remaining items for training.
For easy and fair evaluation, we follow the common strategy in \cite{He:www2017:NCF,Tang:WSDM2018:caser,Kang:ICDM2018:SAS}, pairing each ground truth item in the test set with 100 randomly sampled \textit{negative} items that the user has not interacted with.
To make the sampling reliable and representative~\cite{Huang:sigir2018:ISR}, these 100 negative items are sampled according to their popularity.
Hence, the task becomes to rank these negative items with the ground truth item for each user.

\textbf{Evaluation Metrics}.
To evaluate the ranking list of all the models, we employ a variety of evaluation metrics, including \textit{Hit Ratio} (HR), \textit{Normalized Discounted Cumulative Gain} (NDCG), and \textit{Mean Reciprocal Rank} (MRR).
Considering we only have one ground truth item for each user, HR@$k$ is equivalent to Recall@$k$ and proportional to Precision@$k$; MRR is equivalent to Mean Average Precision (MAP).
In this work, we report HR and NDCG with $k={1,5,10}$.
For all these metrics, the higher the value, the better the performance.

\subsection{Baselines \& Implementation Details}

To verify the effectiveness of our method, we compare it with the following representative baselines:
\begin{itemize}[leftmargin=0.6cm]
    \item \textbf{POP}: It is the simplest baseline that ranks items according to their popularity judged by the number of interactions.
    \item \textbf{BPR-MF}~\cite{Rendle:UAI2009:BBP}: It optimizes the matrix factorization with implicit feedback using a pairwise ranking loss.
    \item \textbf{NCF}~\cite{He:www2017:NCF}: It models user–item interactions with a MLP instead of the inner product in matrix factorization.
    \item \textbf{FPMC}~\cite{Rendle:WWW2010:FPM}: It captures users' general taste as well as their sequential behaviors by combing MF with first-order MCs.
    \item \textbf{GRU4Rec}~\cite{Hidasi:ICLR2016:gru4rec}: It uses GRU with ranking based loss to model user sequences for session based recommendation.
    \item \textbf{GRU4Rec$^{+}$}~\cite{Hidasi:CIKM2018:RNN}: It is an improved version of GRU4Rec with a new class of loss functions and sampling strategy.
    \item \textbf{Caser}~\cite{Tang:WSDM2018:caser}: It employs CNN in both horizontal and vertical way to model high-order MCs for sequential recommendation.
    \item \textbf{SASRec}~\cite{Kang:ICDM2018:SAS}: It uses a left-to-right Transformer language model to capture users' sequential behaviors, and achieves state-of-the-art performance on sequential recommendation.
\end{itemize}

For NCF\footnote{\url{https://github.com/hexiangnan/neural_collaborative_filtering}}, GRU4Rec\footnote{\label{gru4rec}\url{https://github.com/hidasib/GRU4Rec}}, GRU4Rec$^+$\textsuperscript{\ref{gru4rec}}, Caser\footnote{\url{https://github.com/graytowne/caser_pytorch}}, and SASRec\footnote{\url{https://github.com/kang205/SASRec}}, we use code provided by the corresponding authors.
For BPR-MF and FPMC, we implement them using \texttt{TensorFlow}.
For common hyper-parameters in all models, we consider the hidden dimension size $d$ from $\{16, 32, 64, 128, 256\}$, the $\ell_2$ regularizer from \{1, 0.1, 0.01, 0.001, 0.0001\}, and dropout rate from \{$0, 0.1, 0.2,\cdots, 0.9$\}.
All other hyper-parameters (\textit{e.g.}, Markov order in Caser) and initialization strategies are either followed the suggestion from the methods' authors or tuned on the validation sets.
We report the results of each baseline under its optimal hyper-parameter settings.

We implement BERT4Rec\footnote{\url{https://github.com/FeiSun/BERT4Rec}} with \texttt{TensorFlow}.
All parameters are initialized using truncated normal distribution in the range $[-0.02, 0.02]$.
We train the model using Adam~\cite{Kingma:iclr2015:Adam} with learning rate of 1e-4, $\beta_1 = 0.9, \beta_2 = 0.999$, $\ell_2$ weight decay of 0.01, and linear decay of the learning rate.
The gradient is clipped when its $\ell_2$ norm exceeds a threshold of 5. %
For fair comparison, we set the layer number $L=2$ and head number $h=2$ and use the same maximum sequence length as in~\cite{Kang:ICDM2018:SAS}, $N=200$ for ML-1m and ML-20m, $N=50$ for Beauty and Steam datasets. %
For head setting, we empirically set the dimensionality of each head as 32 (single head if $d<32$).
We tune the mask proportion $\rho$ using the validation set, resulting in $\rho=0.6$ for Beauty, $\rho=0.4$ for Steam, and $\rho=0.2$ for ML-1m and ML-20m. %
All the models are trained from scratch without any pre-training on a single NVIDIA GeForce GTX 1080 Ti GPU with a batch size of 256.

\subsection{Overall Performance Comparison}
\begin{table*}[]
\setlength{\tabcolsep}{0.62em}
    \centering
    \caption{Performance comparison of different methods on next-item prediction. Bold scores are the best in each row, while underlined scores are the second best. Improvements over baselines are statistically significant with $p<0.01$.} %
    \begin{adjustbox}{max width=\textwidth}
        \begin{tabular}{l l c c c c c c c c c c}
        \toprule
        Datasets & Metric & POP & BPR-MF & NCF & FPMC & GRU4Rec & GRU4Rec$^+$ & Caser & SASRec & BERT4Rec & Improv.\\
        \midrule
        \multirow{6}{*}{Beauty} & HR@1 & 0.0077 & 0.0415 & 0.0407 & 0.0435 & 0.0402 & 0.0551 & 0.0475 & \underline{0.0906} & \textbf{0.0953} & ~~5.19\% \\
         & HR@5 & 0.0392 & 0.1209 & 0.1305 & 0.1387 & 0.1315 & 0.1781 & 0.1625 & \underline{0.1934} & \textbf{0.2207} & 14.12\%\\
         & HR@10 & 0.0762 & 0.1992 & 0.2142 & 0.2401 & 0.2343 & 0.2654 & 0.2590 & \underline{0.2653} & \textbf{0.3025} & 14.02\%\\
         & NDCG@5 & 0.0230 & 0.0814 & 0.0855 & 0.0902 & 0.0812 & 0.1172 & 0.1050 & \underline{0.1436} & \textbf{0.1599} & 11.35\%\\
         & NDCG@10 & 0.0349 & 0.1064 & 0.1124 & 0.1211 & 0.1074 & 0.1453 & 0.1360 & \underline{0.1633} & \textbf{0.1862} & 14.02\%\\
          & MRR & 0.0437 & 0.1006 & 0.1043 & 0.1056 & 0.1023 & 0.1299 & 0.1205 & \underline{0.1536} & \textbf{0.1701} & 10.74\%\\
         \midrule
        \multirow{6}{*}{Steam} & HR@1 & 0.0159 & 0.0314 & 0.0246 & 0.0358 & 0.0574 & 0.0812 & 0.0495 & \underline{0.0885} & \textbf{0.0957} & ~~8.14\%\\
         & HR@5 & 0.0805 & 0.1177 & 0.1203 & 0.1517 & 0.2171 & 0.2391 & 0.1766 & \underline{0.2559} & \textbf{0.2710} & ~~5.90\%\\
         & HR@10 & 0.1389 & 0.1993 & 0.2169 & 0.2551 & 0.3313 & 0.3594 & 0.2870 & \underline{0.3783} & \textbf{0.4013} & ~~6.08\% \\
         & NDCG@5 & 0.0477 & 0.0744 & 0.0717 & 0.0945 & 0.1370 & 0.1613 & 0.1131 & \underline{0.1727} & \textbf{0.1842} & ~~6.66\% \\
         & NDCG@10 & 0.0665 & 0.1005 & 0.1026 & 0.1283 & 0.1802 & 0.2053 & 0.1484 & \underline{0.2147} & \textbf{0.2261} & ~~5.31\%\\
         & MRR & 0.0669 & 0.0942 & 0.0932 & 0.1139 & 0.1420 & 0.1757 & 0.1305 & \underline{0.1874} & \textbf{0.1949} & ~~4.00\%\\
         \midrule
        \multirow{6}{*}{ML-1m} & HR@1 & 0.0141 & 0.0914 & 0.0397 & 0.1386 & 0.1583 & 0.2092 & 0.2194 & \underline{0.2351} & \textbf{0.2863} & 21.78\%\\
         & HR@5 & 0.0715 & 0.2866 & 0.1932 & 0.4297 & 0.4673 & 0.5103 & 0.5353 & \underline{0.5434} & \textbf{0.5876} & ~~8.13\%\\
         & HR@10 & 0.1358 & 0.4301 & 0.3477 & 0.5946 & 0.6207 & 0.6351 &  \underline{0.6692} & 0.6629 & \textbf{0.6970} & ~~4.15\%\\
         & NDCG@5 & 0.0416 & 0.1903 & 0.1146 & 0.2885 & 0.3196 & 0.3705 & 0.3832 & \underline{0.3980} & \textbf{0.4454} & 11.91\%\\
         & NDCG@10 & 0.0621 & 0.2365 & 0.1640 & 0.3439 & 0.3627 & 	0.4064 & 0.4268 & \underline{0.4368} & \textbf{0.4818} & 10.32\%\\
         & MRR & 0.0627 & 0.2009 & 0.1358 & 0.2891 & 0.3041 & 	0.3462 & 0.3648 & \underline{0.3790} & \textbf{0.4254} & 12.24\%\\
         \midrule
        \multirow{6}{*}{ML-20m} & HR@1 & 0.0221 & 0.0553 & 0.0231 & 0.1079 & 0.1459 & 0.2021 & 0.1232 & \underline{0.2544} & \textbf{0.3440} & 35.22\%\\
         & HR@5 & 0.0805 & 0.2128 & 0.1358 & 0.3601 & 0.4657 & 0.5118 & 0.3804 & \underline{0.5727} & \textbf{0.6323} & 10.41\%\\
         & HR@10 & 0.1378 & 0.3538 & 0.2922 & 0.5201 & 0.5844 & 0.6524 & 0.5427 & \underline{0.7136} & \textbf{0.7473} & ~~4.72\% \\
         & NDCG@5 & 0.0511 & 0.1332 & 0.0771 & 0.2239 & 0.3090 & 0.3630 & 0.2538 & \underline{0.4208} & \textbf{0.4967} & 18.04\%\\
         & NDCG@10 & 0.0695 & 0.1786 & 0.1271 & 0.2895 & 0.3637 & 	0.4087 & 0.3062 & \underline{0.4665} & \textbf{0.5340} & 14.47\% \\
         & MRR & 0.0709 & 0.1503 & 0.1072 & 0.2273 & 0.2967 & 0.3476 & 0.2529 & \underline{0.4026} & \textbf{0.4785} & 18.85\% \\
        \bottomrule
        \end{tabular}
    \end{adjustbox}
    \label{tab:result}
\end{table*}

Table~\ref{tab:result} summarized the best results of all models on four benchmark datasets.
The last column is the improvements of BERT4Rec relative to the best baseline.
We omit the NDCG@1 results since it is equal to HR@1 in our experiments.
It can be observed that:

The non-personalized POP method gives the worst performance\footnote{What needs to be pointed out is that such low scores for POP are because the negative samples are sampled according to the items' popularity.} on all datasets since it does not model user's personalized preference using the historical records.
Among all the baseline methods, sequential methods (\textit{e.g.}, FPMC and GRU4Rec$^+$) outperforms non-sequential methods (\textit{e.g.}, BPR-MF and NCF) on all datasets consistently.
Compared with BPR-MF, the main improvement of FPMC is that it models users' historical records in a sequential way.
This observation verifies that considering sequential information is beneficial for improving performances in recommendation systems.

Among sequential recommendation baselines, Caser outperforms FPMC on all datasets especially for the dense dataset ML-1m, suggesting that high-order MCs is beneficial for sequential recommendation.
However, high-order MCs usually use very small order $L$ since they do not scale well with the order $L$.
This causes Caser to perform worse than GRU4Rec$^+$ and SASRec, especially on sparse datasets.
Furthermore, SASRec performs distinctly better than GRU4Rec and GRU4Rec$^+$, suggesting that self-attention mechanism is a more powerful tool for sequential recommendation.

According to the results, it is obvious that BERT4Rec performs best  among all methods on four datasets in terms of all evaluation metrics.
It gains \textbf{7.24\%} HR@10, \textbf{11.03\%} NDCG@10, and \textbf{11.46\%} MRR improvements (on average) against the strongest baselines.

\noindent \textbf{Question 1}: \textit{Do the gains come from the bidirectional self-attention model or from the Cloze objective?}

To answer this question, we try to isolate the effects of these two factors by constraining the Cloze task to mask only one item at a time.
In this way, the main difference between our BERT4Rec (with 1 mask) and SASRec is that BERT4Rec predicts the target item jointly conditioning on both left and right context.
We report the results on Beauty and ML-1m with $d=256$ in Table~\ref{tab:analysis} due to the space limitation.
The results show that BERT4Rec with 1 mask significantly outperforms SASRec on all metrics.
It demonstrates the importance of bidirectional representations for sequential recommendation.
Besides, the last two rows indicate that the Cloze objective also improves the performances.
Detailed analysis of the mask proportion $\rho$ in Cloze task can be found in \S~\ref{sec:mask}

\begin{table}
    \centering
    \caption{Analysis on bidirection and Cloze with $d=256$.}
    \label{tab:analysis}
    \begin{adjustbox}{max width=\linewidth}
    \begin{tabular}
        {l c c c c c c}
        \toprule
       \multirow{2}{*}{Model} & \multicolumn{3}{c}{Beauty} & \multicolumn{3}{c}{ML-1m} \\ \cmidrule(lr){2-4} \cmidrule(lr){5-7}
        & HR@10 & NDCG@10 & MRR & HR@10 & NDCG@10 & MRR \\ \midrule
        SASRec & 0.2653 & 0.1633 & 0.1536 & 0.6629 & 0.4368 & 0.3790 \\
        BERT4Rec (1 mask) & 0.2940 & 0.1769 & 0.1618 & 0.6869 & 0.4696 & 0.4127 \\
        BERT4Rec & 0.3025 & 0.1862 & 0.1701 & 0.6970 & 0.4818 & 0.4254 \\ 
        \bottomrule
    \end{tabular}
    \end{adjustbox}
\end{table}

\noindent \textbf{Question 2}: \textit{Why and how does bidirectional model outperform unidirectional models?}
\begin{figure}
\vspace{2mm}
    \centering
    \begin{subfigure}[b]{0.45\linewidth}
        \includegraphics[width=\textwidth]{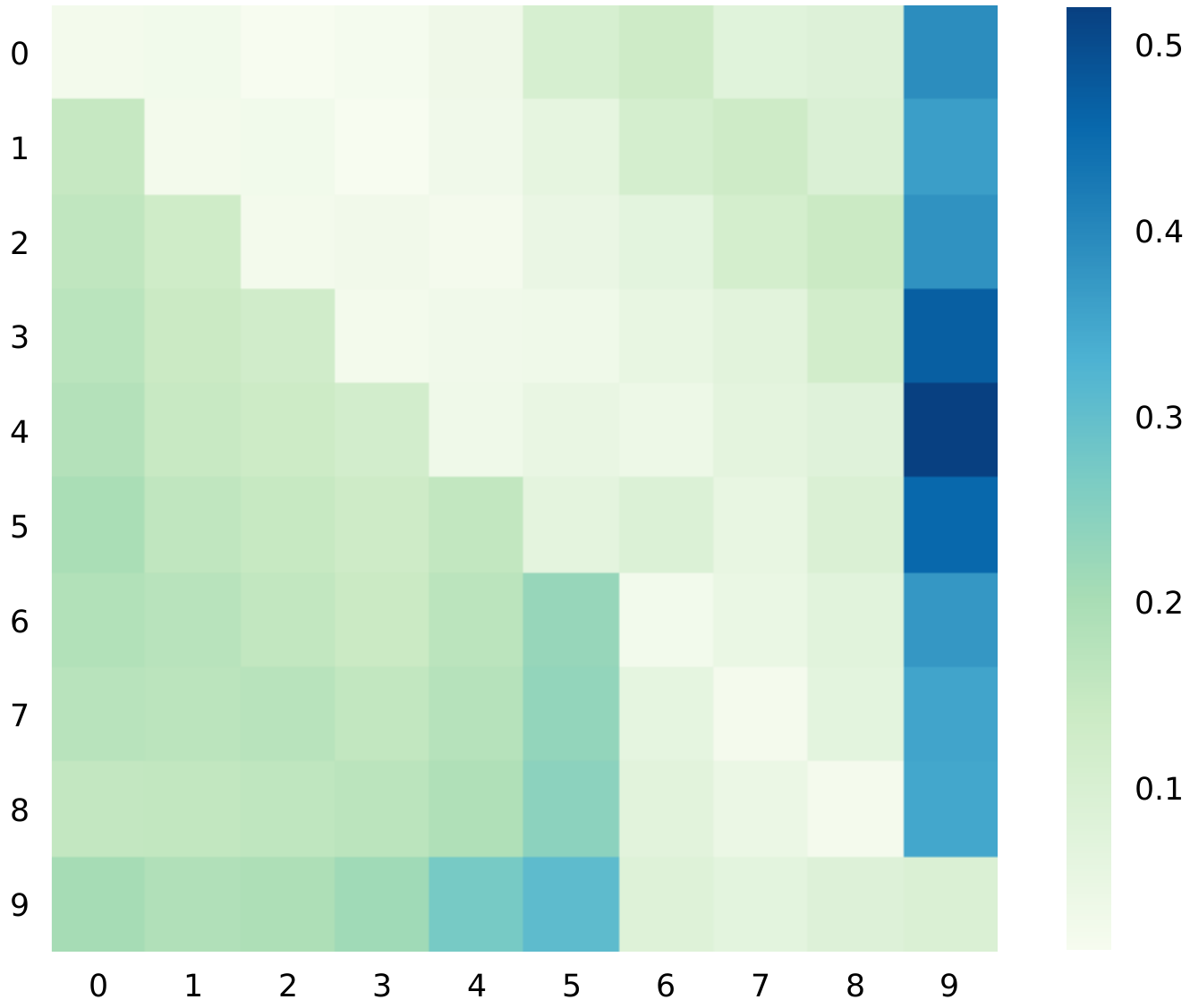}
        \caption{Layer 1, head 1}
        \label{fig:l1h1}
    \end{subfigure}
    \quad
    \begin{subfigure}[b]{0.45\linewidth}
        \includegraphics[width=\textwidth]{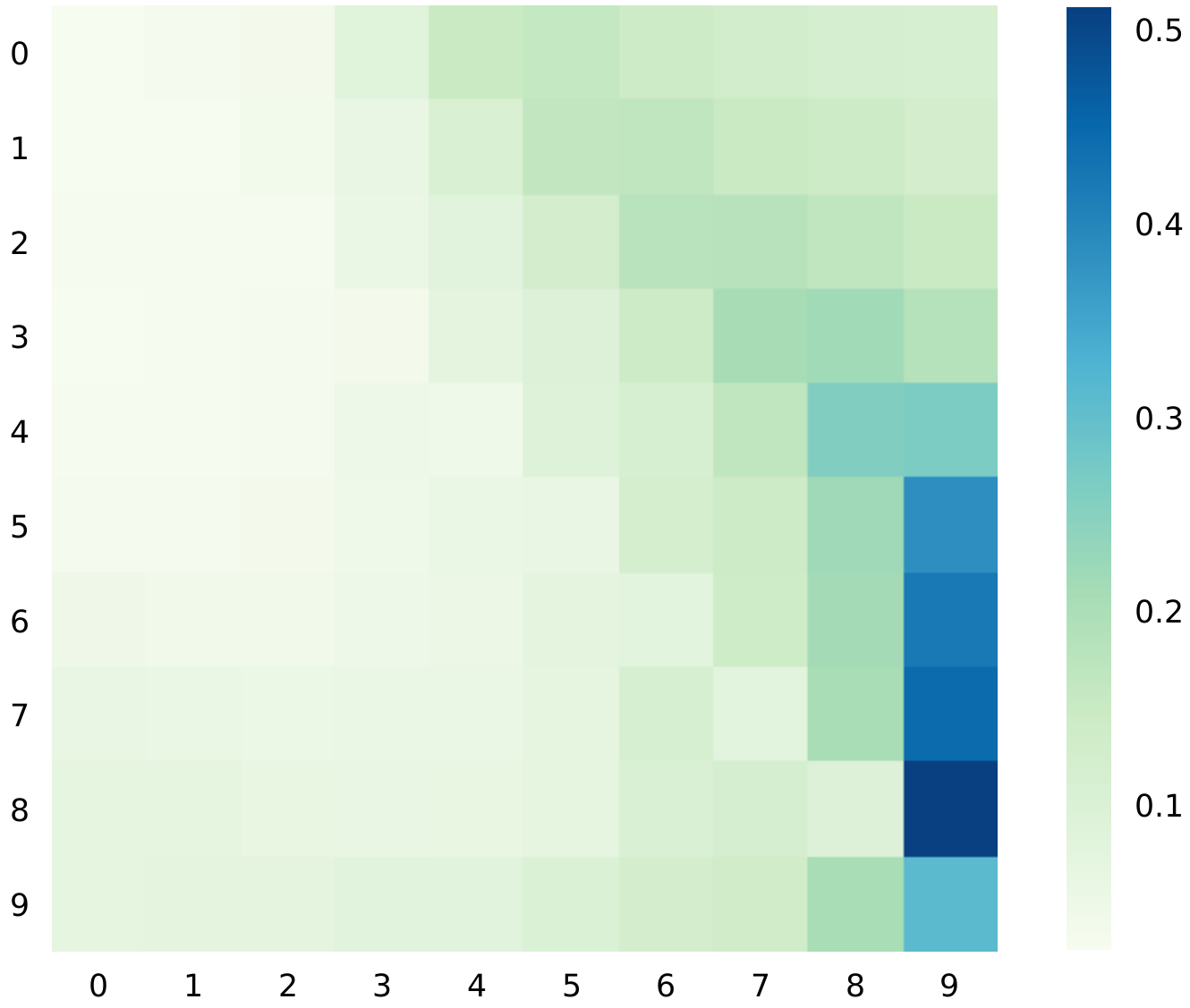}
        \caption{Layer 1, head 2}
        \label{fig:l1h2}
    \end{subfigure}
    \begin{subfigure}[b]{0.47\linewidth}
     ~~~\includegraphics[width=\textwidth]{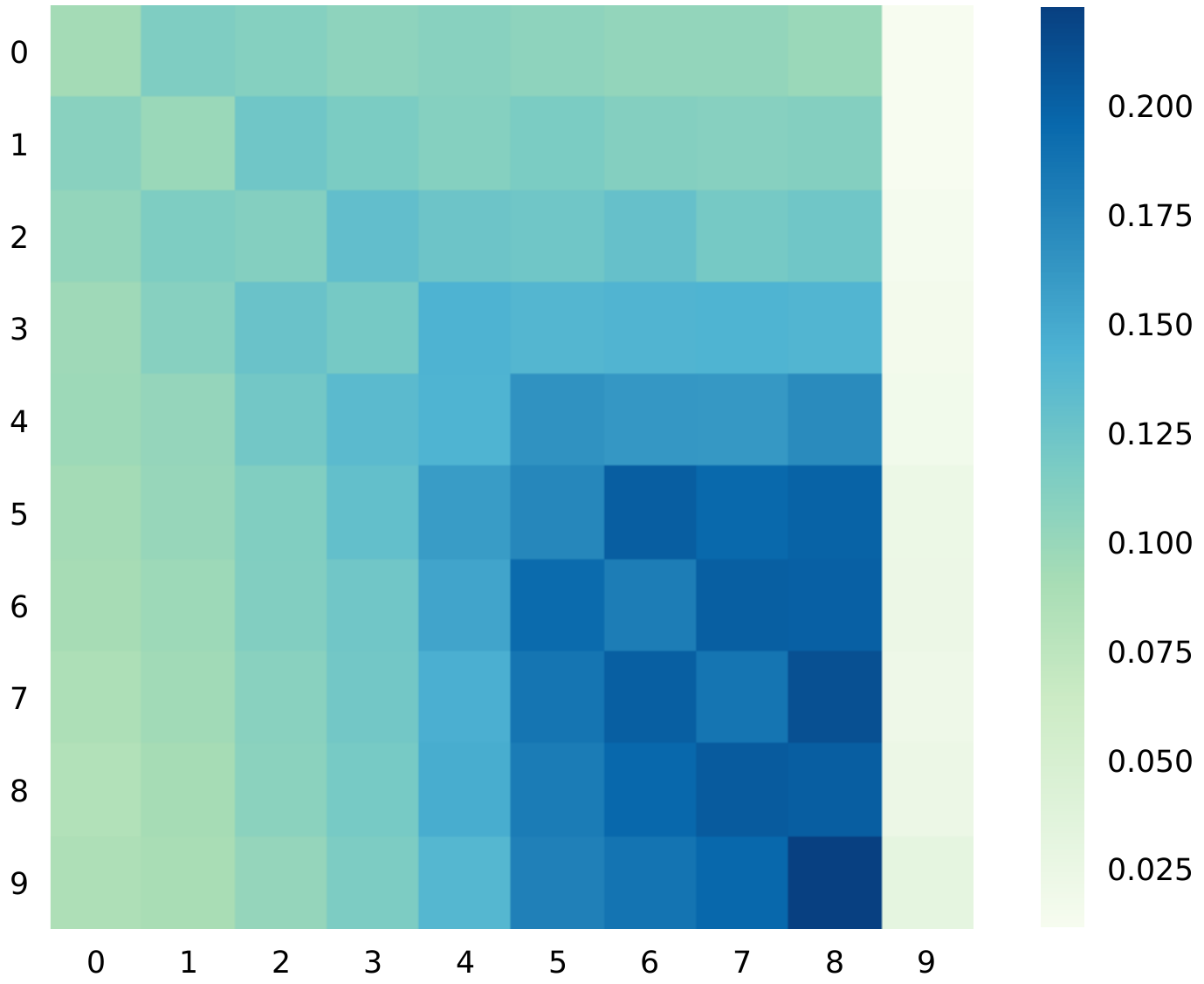}
        \caption{Layer 2, head 2}
        \label{fig:mouse}
    \end{subfigure}
    \quad
    \begin{subfigure}[b]{0.47\linewidth}
        \includegraphics[width=\textwidth]{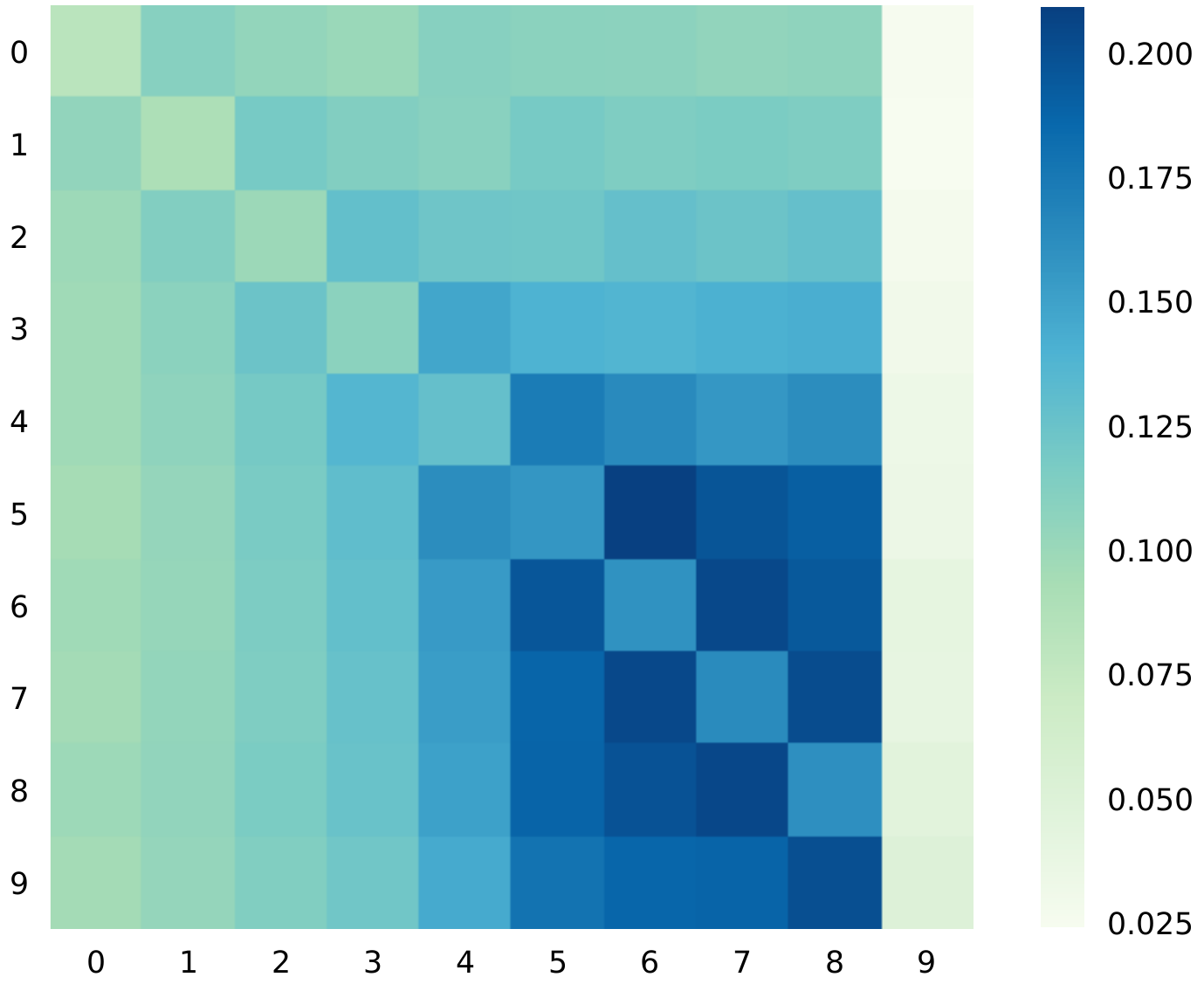}
        \caption{Layer 2, head 4}
        \label{fig:mouse}
    \end{subfigure}
    \caption{Heat-maps of average attention weights on Beauty, the last position ``9'' denotes ``\texttt{[mask]}'' (best viewed in color).}
    \label{fig:att_vis}
\end{figure}

To answer this question, we try to reveal meaningful patterns by visualizing the average attention weights of the last 10 items during the test on Beauty in Figure~\ref{fig:att_vis}. 
Due to the space limitation, we only report four representative attention heat-maps in different layers and heads.

\input{dim_exp_fig}
We make several observations from the results.
\begin {enumerate*}[label=\bfseries\itshape\alph*\upshape)]
\item Attention varies across different heads.
For example, in layer 1, head 1 tends to attend on items at the left side while head 2 prefers to attend on items on the right side.
\item Attention varies across different layers.
Apparently, attentions in layer 2 tend to focus on more recent items.
This is because layer 2 is directly connected to the output layer and the recent items play a more important role in predicting the future.
Another interesting pattern is that heads in Figure~\ref{fig:l1h1} and~\ref{fig:l1h2} also tend to attend on \texttt{[mask]}\footnote{This phenomenon also exists in text sequence modeling using BERT.}.
It may be a way for self-attention to propagate sequence-level state to the item level.
\item Finally and most importantly, unlike unidirectional model can only attend on items at the left side, items in BERT4Rec tend to attend on the items at both sides.
This indicates that bidirectional is essential and beneficial for user behavior sequence modeling.
\end {enumerate*}

In the following studies, we examine the impact of the hyper-parameters, including the hidden dimensionality $d$, the mask proportion $\rho$, and the maximum sequence length $N$.
We analyze one hyper-parameter at a time by fixing the remaining hyper-parameters at their optimal settings. 
Due to space limitation, we only report NDCG@10 and HR@10 for the follow-up experiments. %

\subsection{Impact of Hidden Dimensionality $d$}

We now study how the hidden dimensionality $d$ affects the recommendation performance.
Figure~\ref{fig:dim_exp} shows NDCG@10 and HR@10 for neural sequential methods with the hidden dimensionality $d$ varying from 16 to 256 while keeping other optimal hyper-parameters unchanged.
We make some observations from this figure.

The most obvious observation from these sub-figures is that the performance of each model tends to converge as the dimensionality increases.
A larger hidden dimensionality does not necessarily lead to better model performance, especially on sparse datasets like Beauty and Steam.
This is probably caused by overfitting.
In terms of details, Caser performs unstably on four datasets, which might limit its usefulness.
Self-attention based methods (\textit{i.e.}, SASRec and BERT4Rec) achieve superior performances on all datasets. %
Finally, our model consistently outperforms all other baselines on all datasets even with a relatively small hidden dimensionality.
Considering that our model achieves satisfactory performance with $d {\geq} 64$, we only report the results with $d{=}64$ in the following analysis.

\subsection{Impact of Mask Proportion $\rho$}
\label{sec:mask}

\input{mask_fig}

As described in \S~\ref{sec:ml}, mask proportion $\rho$ is a key factor in model training, which directly affects the loss function (Equation~\ref{eq:loss}).
Obviously, mask proportion $\rho$ should not be too small or it is not enough to learn a strong model.
Meanwhile, it should not be too large, otherwise, it would be hard to train since there are too many items to guess based on a few contexts in such case.
To examine this, we study how mask proportion $\rho$ affects the recommendation performances on different datasets.

Figure~\ref{fig:mask_fig} shows the results with varying mask proportion $\rho$ from 0.1 to 0.9.
Considering the results with $\rho > 0.6$ on all datasets, a general pattern emerges, the performances decreasing as $\rho$ increases.
From the results of the first two columns, it is easy to see that $\rho=0.2$ performs better than $\rho=0.1$ on all datasets.
These results verify what we claimed above.

In addition, we observe that the optimal $\rho$ is highly dependent on the sequence length of the dataset.
For the datasets with short sequence length (\textit{e.g.}, Beauty and Steam), the best performances are achieved at $\rho{=}0.6$ (Beauty) and $\rho{=}0.4$ (Steam), while the datasets with long sequence length (\textit{e.g.}, ML-1m and ML-20m) prefer a small $\rho{=}0.2$.
This is reasonable since, compared with short sequence datasets, a large $\rho$ in long sequence datasets means much more items that need to be predicted. %
Take ML-1m and Beauty as example, $\rho{=}0.6$ means we need to predict $98 {=} \lfloor 163.5{\times} 0.6 \rfloor $ items on average per sequence for ML-1m, while it is only $5 {=} \lfloor 8.8{\times} 0.6 \rfloor$ items for Beauty.
The former is too hard for model training.

\subsection{Impact of Maximum Sequence Length $N$}

\begin{table}[t]
\renewcommand{\arraystretch}{1}
    \centering
    \caption{Performance with different maximum length $N$.}
    \label{tab:seq}
    \begin{adjustbox}{max width=\linewidth}
        \begin{tabular}{l l r r r r r }
        \toprule
         & & \multicolumn{1}{c}{10} & \multicolumn{1}{c}{20} & \multicolumn{1}{c}{30} & \multicolumn{1}{c}{40}  & \multicolumn{1}{c}{50}  \\ \midrule
       \multirow{3}{*}{Beauty} & \#samples/s & 5504 & 3256 & 2284 &  1776 & 1441\\
       & HR@10 & 0.3006 & 0.3061 & 0.3057 & 0.3054 & 0.3047 \\
       & NDCG@10 & 0.1826 & 0.1875 & 0.1837 & 0.1833 & 0.1832 \\
        \bottomrule \toprule
       &  & \multicolumn{1}{c}{10} & \multicolumn{1}{c}{50} & \multicolumn{1}{c}{100} & \multicolumn{1}{c}{200}  & \multicolumn{1}{c}{400}  \\ \midrule
       \multirow{3}{*}{ML-1m} & \#samples/s & 14255 & 8890 & 5711 & 2918 & 1213\\
       & HR@10 & 0.6788 & 0.6854 & 0.6947 & 0.6955 & 0.6898 \\
       & NDCG@10 & 0.4631 & 0.4743 & 0.4758 & 0.4759 & 0.4715 \\
        \bottomrule
    \end{tabular}
    \end{adjustbox}
\end{table}

We also investigate the effect of the maximum sequence length $N$ on model's recommendation performances and efficiency.

Table~\ref{tab:seq} shows recommendation performances and training speed with different maximum length $N$ on Beauty and ML-1m.
We observe that the proper maximum length $N$ is also highly dependent on the average sequence length of the dataset.
Beauty prefers a smaller $N=20$, while ML-1m achieves the best performances on $N=200$.
This indicates that a user's behavior is affected by more recent items on short sequence datasets and less recent items for long sequence datasets.
The model does not consistently benefit from a larger $N$ since a larger $N$ tends to introduce both extra information and more noise.
However, our model performs very stably as the length $N$ becomes larger.
This indicates that our model can attend to the informative items from the noisy historical records.

A scalability concern about BERT4Rec is that its computational complexity per layer is $\mathcal{O}(n^2d)$, quadratic with the length $n$.
Fortunately, the results in Table~\ref{tab:seq} shows that the self-attention layer can be effectively parallelized using GPUs.

\subsection{Ablation Study}

Finally, we perform ablation experiments over a number of key components of BERT4Rec in order to better understand their impacts, including positional embedding (PE), position-wise feed-forward network (PFFN), layer normalization (LN), residual connection (RC), dropout, the layer number $L$ of self-attention, and the number of heads $h$ in multi-head attention. %
Table~\ref{tab:ablation} shows the results of our default version ($L=2, h=2$) and its eleven variants on all four datasets with dimensionality $d=64$ while keeping other hyper-parameters (\textit{e.g.}, $\rho$) at their optimal settings. %

We introduce the variants and analyze their effects respectively:
\begin{enumerate}%
  \item \textbf{PE}. The results show that removing positional embeddings causes BERT4Rec's performances decreasing dramatically on long sequence datasets (\textit{i.e.}, ML-1m and ML-20m). Without the positional embeddings, the hidden representation $\bm{H}_i^L$ for each item $v_i$ depends only on item embeddings.
  In this situation, we predict different target items using the same hidden representation of ``\texttt{[mask]}''. This makes the model ill-posed. This issue is more serious on long sequence datasets since they have more masked items to predict.
  \item \textbf{PFFN}. The results show that long sequence datasets (\textit{e.g.}, ML-20m) benefit more from PFFN. 
  This is reasonable since a purpose of PFFN is to integrate information from many heads which are preferred by long sequence datasets as discussed in the analysis about head number $h$ in ablation study (\ref{itm:head}). %
  \item \textbf{LN}, \textbf{RC}, and \textbf{Dropout}. These components are introduced mainly to alleviate overfitting. 
  Obviously, they are more effective on small datasets like Beauty.
   To verify their effectiveness on large datasets, we conduct an experiment on ML-20m with layer $L{=}4$.
   The results show that NDCG@10 decreases about 10\% w/o RC.
  \item \textbf{Number of layers $L$}. The results show that stacking Transformer layer can boost performances especially on large datasets (\textit{e.g}, ML-20m).
  This verifies that it is helpful to learn more complex item transition patterns via deep self-attention architecture.
  The decline in Beauty with $L=4$ is largely due to overfitting.
  \item\label{itm:head} \textbf{Head number $h$}. We observe that long sequence datasets (\textit{e.g.}, ML-20m) benefit from a larger $h$ while short sequence datasets (\textit{e.g.}, Beauty) prefer a smaller $h$. This phenomenon is consistent with the empirical result in~\cite{Tang:EMNLP2018:Why} that large $h$ is essential for capturing long distance dependencies with multi-head self-attention.
\end{enumerate}
\begin{table}[t]
\renewcommand{\arraystretch}{1}
\setlength{\tabcolsep}{0.75em}
    \centering
    \caption{Ablation analysis (NDCG@10) on four datasets. Bold score indicates performance better than the default version, while $\downarrow$ indicates performance drop more than 10\%.}
    \label{tab:ablation}
    \begin{adjustbox}{max width=\linewidth}
        \begin{tabular}{l r r r r}
        \toprule
     \multirow{2}{*}{Architecture} & \multicolumn{4}{c}{Dataset} \\ 
         \cmidrule(lr){2-5}
          & \multicolumn{1}{c}{Beauty} & \multicolumn{1}{c}{Steam} & \multicolumn{1}{c}{ML-1m} & \multicolumn{1}{c}{ML-20m} \\ \midrule
        $L=2$, $h=2$  & 0.1832 & 0.2241 & 0.4759 & 0.4513 \\ \midrule
        w/o PE  & 0.1741 & 0.2060 & 0.2155$\downarrow$ & 0.2867$\downarrow$ \\
        w/o PFFN  & 0.1803 & 0.2137 & 0.4544 & 0.4296 \\ \midrule
        w/o LN  & 0.1642$\downarrow$ & 0.2058 & 0.4334 & 0.4186 \\
        w/o RC  & 0.1619$\downarrow$ & 0.2193 & 0.4643 & 0.4483 \\
        w/o Dropout  & 0.1658 & 0.2185 & 0.4553 & 0.4471 \\ \midrule
        1 layer \hspace{\fill}($L=1$) & 0.1782 & 0.2122 & 0.4412 & 0.4238 \\
        3 layers \hspace{\fill} ($L=3$) & \textbf{0.1859} & \textbf{0.2262} & \textbf{0.4864} & \textbf{0.4661}\\
        4 layers \hspace{\fill} ($L=4$) & \textbf{0.1834} & \textbf{0.2279} & \textbf{0.4898} & \textbf{0.4732}\\ \midrule
        1 head \hspace{\fill} ($h=1$) & \textbf{0.1853} & 0.2187 & 0.4568 & 0.4402 \\
        4 heads \hspace{\fill} ($h=4$) & 0.1830 & \textbf{0.2245} & \textbf{0.4770} & \textbf{0.4520} \\
        8 heads \hspace{\fill} ($h=8$) & 0.1823 & \textbf{0.2248} & 0.4743 & \textbf{0.4550} \\
        \bottomrule
    \end{tabular}
    \end{adjustbox}
\end{table}

%% file: dim_exp_fig.tex
\pgfplotsset{
axis background/.style={fill=gallery},
grid=both,
  xtick pos=left,
  ytick pos=left,
  tick style={
    major grid style={style=white,line width=1pt},
    minor grid style=bgc,
    draw=none
    },
  minor tick num=1,
  ymajorgrids,
	major grid style={draw=white},
	y axis line style={opacity=0},
	tickwidth=0pt,
}

\begin{figure*}[]
\centering
    \begin{tikzpicture}[scale=0.5]
	\begin{groupplot}[
	    group style={group size=4 by 2,
	        horizontal sep = 42pt}, 
	    width=0.5\textwidth,
	    height=0.4\textwidth,
	    xlabel=\large Dimensionality,
        ylabel=NDCG@10,
        xticklabels={16, 32, 64, 128, 256},
        xtick={1,2,3,4,5},
        ymajorgrids,
        major grid style={draw=white},
        y axis line style={opacity=0},
        tickwidth=0pt,
        yticklabel style={
        /pgf/number format/fixed,
        /pgf/number format/precision=5
        },
        scaled y ticks=false,
        every axis title/.append style={at={(0.1,0.8)},font=\bfseries}
	    ]
		\nextgroupplot[
		legend style = {
		  font=\small,
          draw=none, 
          fill=none,
          column sep = 1pt, 
          /tikz/every even column/.append style={column sep=5mm},
          legend columns = -1, 
          legend to name = grouplegend},
		title=Beauty, 
		]
		
		\addplot[thick,color=tuatara,mark=pentagon] coordinates {
          (1, 0.0722)
          (2, 0.0869)
          (3, 0.1041)
          (4, 0.1053)
          (5, 0.1074)
        }; \addlegendentry{GRU4Rec}
         \addplot[thick,color=free_speech_aquamarine,mark=triangle*] coordinates {
          (1, 0.0977)
          (2, 0.1113)
          (3, 0.1311)
          (4, 0.1435)
          (5, 0.1453)
        }; \addlegendentry{GRU4Rec$^{+}$}
        \addplot[thick,color=matisse,mark=square] coordinates {
          (1, 0.0906)
          (2, 0.1091)
          (3, 0.1230)
          (4, 0.1353)
          (5, 0.1360)
        };\addlegendentry{Caser}
        \addplot[thick,color=sun_shade,mark=diamond] coordinates {
          (1, 0.1131)
          (2, 0.1398)
          (3, 0.1575)
          (4, 0.1639)
          (5, 0.1633)
        };\addlegendentry{SASRec}
        \addplot[thick,color=flamingo,mark=*] coordinates {
          (1, 0.1325)
          (2, 0.1668)
          (3, 0.1832)
          (4, 0.1833)
          (5, 0.1862)
        };\addlegendentry{BERT4Rec}
        
        \nextgroupplot[
        title=Steam]
		\addplot[thick,color=tuatara,mark=pentagon] coordinates {
          (1, 0.1141)
          (2, 0.1421)
          (3, 0.1686)
          (4, 0.1754)
          (5, 0.1802)
        };
         \addplot[thick,color=free_speech_aquamarine,mark=triangle*] coordinates {
          (1, 0.1361)
          (2, 0.1780)
          (3, 0.1979)
          (4, 0.2031)
          (5, 0.2053)
        };
        \addplot[thick,color=matisse,mark=square] coordinates {
          (1, 0.1225)
          (2, 0.1364)
          (3, 0.1453)
          (4, 0.1484)
          (5, 0.1477)
        };
        \addplot[thick,color=sun_shade,mark=diamond] coordinates {
          (1, 0.1428)
          (2, 0.1874)
          (3, 0.2088)
          (4, 0.2123)
          (5, 0.2147)
        };
        \addplot[thick,color=flamingo,mark=*] coordinates {
          (1, 0.1890)
          (2, 0.2135)
          (3, 0.2241)
          (4, 0.2261)
          (5, 0.2249)
        };
        
        \nextgroupplot[title=ML-1m]
		\addplot[thick,color=tuatara,mark=pentagon] coordinates {
          (1, 0.2504)
          (2, 0.2826)
          (3, 0.3319)
          (4, 0.3573)
          (5, 0.3627)
        };
         \addplot[thick,color=free_speech_aquamarine,mark=triangle*] coordinates {
          (1, 0.2944)
          (2, 0.3379)
          (3, 0.3851)
          (4, 0.3997)
          (5, 0.4064)
        };
        \addplot[thick,color=matisse,mark=square] coordinates {
          (1, 0.3587)
          (2, 0.3921)
          (3, 0.4198)
          (4, 0.4268)
          (5, 0.4243)
        };
        \addplot[thick,color=sun_shade,mark=diamond] coordinates {
          (1, 0.3080)
          (2, 0.3941)
          (3, 0.4238)
          (4, 0.4327)
          (5, 0.4368)
        };
        \addplot[thick,color=flamingo,mark=*] coordinates {
          (1, 0.3718)
          (2, 0.4327)
          (3, 0.4759)
          (4, 0.4787)
          (5, 0.4818)
        };
        
        \nextgroupplot[title=ML-20m]
		\addplot[thick,color=tuatara,mark=pentagon] coordinates {
          (1, 0.1900)
          (2, 0.2506)
          (3, 0.3134)
          (4, 0.3495)
          (5, 0.3637)
        };
         \addplot[thick,color=free_speech_aquamarine,mark=triangle*] coordinates {
          (1, 0.2891)
          (2, 0.3506)
          (3, 0.3743)
          (4, 0.3975)
          (5, 0.4087)
        };
        \addplot[thick,color=matisse,mark=square] coordinates {
          (1, 0.2302)
          (2, 0.2605)
          (3, 0.2936)
          (4, 0.3044)
          (5, 0.3062)
        };
        \addplot[thick,color=sun_shade,mark=diamond] coordinates {
          (1, 0.2535)
          (2, 0.3177)
          (3, 0.3984)
          (4, 0.4390)
          (5, 0.4665)
        };
        \addplot[thick,color=flamingo,mark=*] coordinates {
          (1, 0.3261)
          (2, 0.3900)
          (3, 0.4513)
          (4, 0.5018)
          (5, 0.5340)
        };

        \nextgroupplot[
        title=Beauty,
        ylabel=HR@10]
		\addplot[thick,color=tuatara,mark=pentagon] coordinates {
          (1, 0.1549)
          (2, 0.1907)
          (3, 0.2267)
          (4, 0.2307)
          (5, 0.2343)
        };
         \addplot[thick,color=free_speech_aquamarine,mark=triangle*] coordinates {
          (1, 0.1965)
          (2, 0.2200)
          (3, 0.2508)
          (4, 0.2631)
          (5, 0.2654)
        };
        \addplot[thick,color=matisse,mark=square] coordinates {
          (1, 0.1898)
          (2, 0.2203)
          (3, 0.2441)
          (4, 0.2580)
          (5, 0.2590)
        };
        \addplot[thick,color=sun_shade,mark=diamond] coordinates {
          (1, 0.2264)
          (2, 0.2567)
          (3, 0.2662)
          (4, 0.2672)
          (5, 0.2653)
        };
        \addplot[thick,color=flamingo,mark=*] coordinates {
          (1, 0.2488)
          (2, 0.2839)
          (3, 0.3047)
          (4, 0.3034)
          (5, 0.3025)
        };
        
        \nextgroupplot[
        title=Steam,
        ylabel=HR@10]
		\addplot[thick,color=tuatara,mark=pentagon] coordinates {
          (1, 0.2395)
          (2, 0.2778)
          (3, 0.3235)
          (4, 0.3291)
          (5, 0.3313)
        };
         \addplot[thick,color=free_speech_aquamarine,mark=triangle*] coordinates {
          (1, 0.2848)
          (2, 0.3278)
          (3, 0.3501)
          (4, 0.3559)
          (5, 0.3594)
        };
        \addplot[thick,color=matisse,mark=square] coordinates {
          (1, 0.2482)
          (2, 0.2696)
          (3, 0.2834)
          (4, 0.2870)
          (5, 0.2857)
        };
        \addplot[thick,color=sun_shade,mark=diamond] coordinates {
          (1, 0.2897)
          (2, 0.3492)
          (3, 0.3737)
          (4, 0.3752)
          (5, 0.3783)
        };
        \addplot[thick,color=flamingo,mark=*] coordinates {
          (1, 0.3534)
          (2, 0.3882)
          (3, 0.4007)
          (4, 0.4013)
          (5, 0.3998)
        };
        
        \nextgroupplot[
        title=ML-1m,
        ylabel=HR@10]
		\addplot[thick,color=tuatara,mark=pentagon] coordinates {
          (1, 0.4811)
          (2, 0.5293)
          (3, 0.5825)
          (4, 0.6126)
          (5, 0.6207)
        };
         \addplot[thick,color=free_speech_aquamarine,mark=triangle*] coordinates {
          (1, 0.5233)
          (2, 0.5684)
          (3, 0.6184)
          (4, 0.6281)
          (5, 0.6351)
        };
        \addplot[thick,color=matisse,mark=square] coordinates {
          (1, 0.5966)
          (2, 0.6419)
          (3, 0.6611)
          (4, 0.6692)
          (5, 0.6639)
        };
        \addplot[thick,color=sun_shade,mark=diamond] coordinates {
          (1, 0.5556)
          (2, 0.6493)
          (3, 0.6720)
          (4, 0.6677)
          (5, 0.6629)
        };
        \addplot[thick,color=flamingo,mark=*] coordinates {
          (1, 0.6060)
          (2, 0.6609)
          (3, 0.6955)
          (4, 0.6997)
          (5, 0.6970)
        };
        
        \nextgroupplot[
        title=ML-20m,
        ylabel=HR@10]
		\addplot[thick,color=tuatara,mark=pentagon] coordinates {
          (1, 0.3990)
          (2, 0.4547)
          (3, 0.5114)
          (4, 0.5667)
          (5, 0.5844)
        };
         \addplot[thick,color=free_speech_aquamarine,mark=triangle*] coordinates {
          (1, 0.5257)
          (2, 0.6028)
          (3, 0.6280)
          (4, 0.6484)
          (5, 0.6524)
        };
        \addplot[thick,color=matisse,mark=square] coordinates {
          (1, 0.4415)
          (2, 0.4922)
          (3, 0.5295)
          (4, 0.5399)
          (5, 0.5427)
        };
        \addplot[thick,color=sun_shade,mark=diamond] coordinates {
          (1, 0.4953)
          (2, 0.5819)
          (3, 0.6657)
          (4, 0.6933)
          (5, 0.7136)
        };
        \addplot[thick,color=flamingo,mark=*] coordinates {
          (1, 0.5593)
          (2, 0.6254)
          (3, 0.6879)
          (4, 0.7230)
          (5, 0.7473)
        };
        
	\end{groupplot}
\node at ($(group c1r1) + (370pt, 95pt)$) {\ref{grouplegend}};
\end{tikzpicture}

    \caption{Effect of the hidden dimensionality $d$ on HR@10 and NDCG@10 for neural sequential models.}
    \label{fig:dim_exp}
\end{figure*}
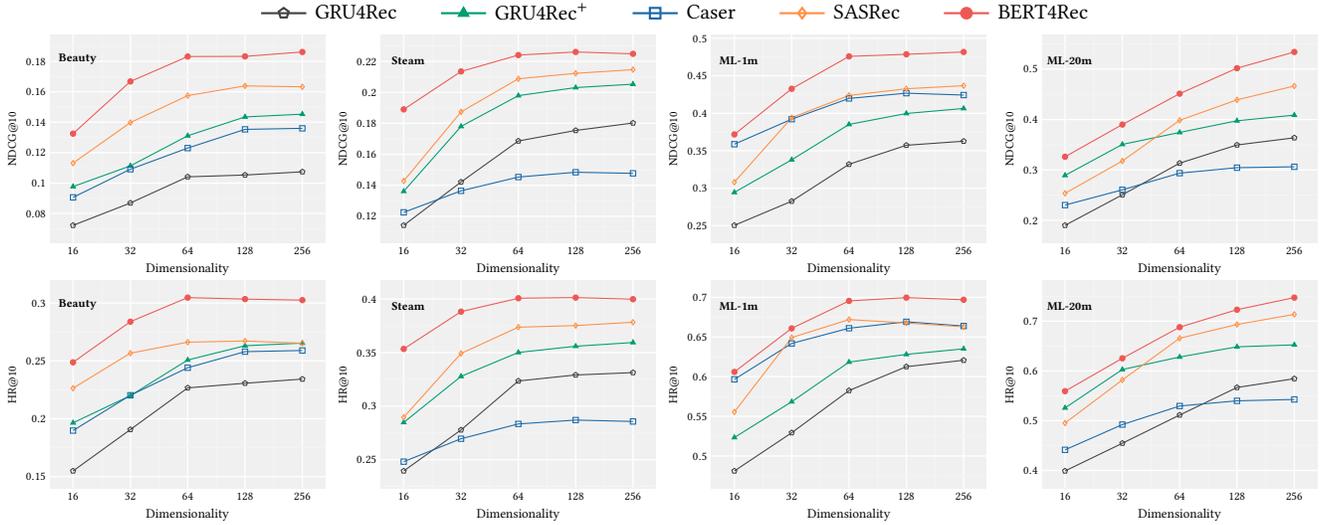

  

%% file: mask_fig.tex
\pgfplotsset{
axis background/.style={fill=gallery},
grid=both,
  xtick pos=left,
  ytick pos=left,
  tick style={
    major grid style={style=white,line width=1pt},
    minor grid style=bgc,
    draw=none
    },
  minor tick num=1,
  ymajorgrids,
	major grid style={draw=white},
	y axis line style={opacity=0},
	tickwidth=0pt,
}

\begin{figure}[ht]
\centering
    \begin{tikzpicture}[scale=0.5]
	\begin{groupplot}[
	    group style={group size=2 by 1,
	        horizontal sep = 48pt}, 
	    xlabel=\Large Dimensionality,
        ylabel=\large HR@10,
        xticklabels={0.1, 0.2, 0.3, 0.4, 0.5, 0.6, 0.7, 0.8, 0.9},
        xtick={1,2,3,4,5,6,7,8,9},
        ymajorgrids,
        major grid style={draw=white},
        y axis line style={opacity=0},
        tickwidth=0pt,
        yticklabel style={
        /pgf/number format/fixed,
        /pgf/number format/precision=5
        },
        scaled y ticks=false,
        every axis title/.append style={at={(0.1,0.8)},font=\bfseries}
	    ]
		\nextgroupplot[
		legend style = {
		  font=\small,
          draw=none, 
          fill=none,
          column sep = 1pt, 
          /tikz/every even column/.append style={column sep=5mm},
          legend columns = -1, 
          legend to name = grouplegend},
		]
		
		\addplot[thick,color=tuatara,mark=*] coordinates {
          (1, 0.2797)
          (2, 0.2869)
          (3, 0.2952)
          (4, 0.2996)
          (5, 0.3028)
          (6, 0.3047)
          (7, 0.3009)
          (8, 0.2968)
          (9, 0.2265)
        }; \addlegendentry{Beauty}
         \addplot[thick,color=free_speech_aquamarine,mark=diamond*] coordinates {
          (1, 0.3831)
          (2, 0.3962)
          (3, 0.3985)
          (4, 0.4007)
          (5, 0.3961)
          (6, 0.3911)
          (7, 0.3884)
          (8, 0.3802)
          (9, 0.3569)
        }; \addlegendentry{Steam}
        \addplot[thick,color=matisse,mark=square] coordinates {
          (1, 0.6877)
          (2, 0.6955)
          (3, 0.6937)
          (4, 0.6875)
          (5, 0.6844)
          (6, 0.6794)
          (7, 0.6745)
          (8, 0.6511)
          (9, 0.6293)
        };\addlegendentry{ML-1m}
        \addplot[thick,color=sun_shade,mark=pentagon] coordinates {
          (1, 0.6781)
          (2, 0.6879)
          (3, 0.6825)
          (4, 0.6822)
          (5, 0.6810)
          (6, 0.6791)
          (7, 0.6667)
          (8, 0.6462)
          (9, 0.6167)
        };\addlegendentry{ML-20m}
        \addplot[mark size=3.6pt, color=tuatara, mark=*,
        ] coordinates {
          (6, 0.3047)
        }; 
         \addplot[mark size=4.2pt, color=free_speech_aquamarine, mark=diamond*] coordinates {
          (4, 0.4007)
        }; 
        \addplot[line width=0.6mm, mark size=3.2pt, color=matisse,mark=square] coordinates {
          (2, 0.6955)
        };
        \addplot[line width=0.6mm, mark size=3.2pt, color=sun_shade,mark=pentagon] coordinates {
          (2, 0.6879)
        };

        \nextgroupplot[
        ylabel=\large NDCG@10]
		\addplot[thick,color=tuatara,mark=*] coordinates {
          (1, 0.1626)
          (2, 0.1693)
          (3, 0.1744)
          (4, 0.1799)
          (5, 0.1809)
          (6, 0.1832)
          (7, 0.1772)
          (8, 0.1749)
          (9, 0.1264)
        };
         \addplot[thick,color=free_speech_aquamarine,mark=diamond*] coordinates {
          (1, 0.2123)
          (2, 0.2206)
          (3, 0.2226)
          (4, 0.2241)
          (5, 0.2207)
          (6, 0.2190)
          (7, 0.2149)
          (8, 0.2096)
          (9, 0.1909)
        }; 
        \addplot[thick,color=matisse,mark=square] coordinates {
          (1, 0.4673)
          (2, 0.4759)
          (3, 0.4743)
          (4, 0.4688)
          (5, 0.4650)
          (6, 0.4587)
          (7, 0.4506)
          (8, 0.4250)
          (9, 0.3942)
        };
        \addplot[thick,color=sun_shade,mark=pentagon] coordinates {
          (1, 0.4444)
          (2, 0.4513)
          (3, 0.4487)
          (4, 0.4471)
          (5, 0.4452)
          (6, 0.4434)
          (7, 0.4292)
          (8, 0.4049)
          (9, 0.3741)
        };
        \addplot[mark size=3.6pt, color=tuatara, mark=*] coordinates {
          (6, 0.1832)
        }; 
         \addplot[mark size=4.2pt, color=free_speech_aquamarine, mark=diamond*] coordinates {
          (4, 0.2241)
        }; 
        \addplot[line width=0.6mm, mark size=3.2pt, color=matisse,mark=square] coordinates {
          (2, 0.4759)
        };
        \addplot[line width=0.6mm, mark size=3.2pt, color=sun_shade,mark=pentagon] coordinates {
          (2, 0.4513)
        };
        
	\end{groupplot}
\node at ($(group c1r1) + (120pt, 95pt)$) {\ref{grouplegend}};
\end{tikzpicture}

    \caption{Performance with different mask proportion $\rho$ on $d=64$. Bold symbols denote the best scores in each line.}
    \label{fig:mask_fig}
\end{figure}
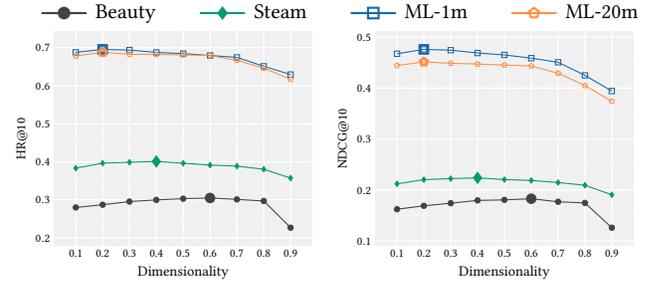

  